\documentclass[12pt,twoside,a4paper]{article}
\usepackage[dvips]{epsfig}
\newcommand{\al}{\alpha}

\newcommand{\g}{\gamma}

\newcommand{\e}{\epsilon}

\newcommand{\si}{\sigma}

\newcommand{\simgt}{\,\rlap{\lower 3.5 pt \hbox{$\mathchar \sim$}} \raise 1pt
 \hbox {$>$}\,}
\newcommand{\simlt}{\,\rlap{\lower 3.5 pt \hbox{$\mathchar \sim$}} \raise 1pt
 \hbox {$<$}\,}

\newcommand{\equ}[2]{\begin{equation} \label{#1} #2 \end{equation} }
\newtheorem{figgur}{Figure}[section]

\begin{document}
\thispagestyle{empty}
\title{\vskip-3cm{\baselineskip14pt
\centerline{\normalsize \hfill MPI--PhT 99-13}
\centerline{\normalsize \hfill February 1999}
\centerline{\normalsize \hfill hep--ph/9903470}
}
\vskip1.5cm Probing the parton densities of virtual photons \\
  with the reaction $\g^*\g\to$jets at LEP 
  \author{B.~P\"otter \\ Max-Planck-Institut f\"ur Physik, \\ F\"ohringer 
  Ring 6, 80805 M\"unchen, Germany. \\ E-mail: poetter@mppmu.mpg.de} }
\date{}
\maketitle

\begin{center}
{\bf Abstract}
\end{center}

\begin{quote}
We present a next-to-leading order calculation of jet production in
$\g^*\g$ collisions from $e^+e^-$ scattering in a region where the
virtuality $Q^2$ of the probing virtual photon is small compared to the 
transverse jet energy. The calculation is based on the phase-space
slicing method. The initial state singularity of the virtual photon is
factorized into the structure function of the virtual photon, using the 
$\overline{\mbox{MS}}$ factorization scheme for virtual photons. 
Numerical results are presented for LEP2 conditions. The perturbative
stability of the pure direct virtual photon approach is compared to
that of including resolved virtual photons in different regions of
$Q^2$. We make predictions for cross sections which suggest that
different parametrizations of virtual photon parton densities should be
distinguishable by measurements of jet cross sections at LEP.
\end{quote}

\newpage

\section{Introduction}

Considerable progress has recently been made in investigating the 
structure of the virtual photon in jet production from $eP$-scattering
at HERA. On the theoretical side calculations are available in leading
order (LO) \cite{1} and next-to-leading order (NLO)
\cite{2,3}. An increasing number of experimental data becomes
available \cite{4} and the confrontation of these 
results with the NLO calculations indicates that the concept of a
resolved virtual photon, i.e., the virtual photon as a source of quarks and
gluons, is necessary to describe the data in the low $Q^2$ region, where 
$Q^2$ is the virtuality of the photon. The parton distribution
function (PDF) of the virtual photon is constructed in
analogy to that of the real photon, with the difference that the 
virtual photon PDF has an extra $Q^2$-dependence built in, in addition
to the usual factorization scale dependence. In the limiting case
$Q^2\to 0$, the virtual photon PDF's reproduce the real photon
PDF's. Since so far only limited data exist on the structure of the
virtual photon \cite{9}, the modeling of the $Q^2$-behaviour of the
virtual photon PDF is still rather ambiguous. Two LO parametrizations
of the virtual photon PDF are available that fit the data 
\cite{9}, namely those of Gl\"uck, Reya and Stratmann \cite{10}, which
have very recently been updated by Gl\"uck, Reya and Schienbein \cite{10b}
(GRS), and those of Schuler and Sj\"ostrand \cite{11} (SaS). The GRS
group has also calculated the virtual photon PDF's in NLO, but no
parametrization is available up to now, since the differences between
the LO and NLO parametrizations are small and the available data is
not yet very precise.

\begin{figure}
\unitlength1mm
\begin{picture}(161,45)
\put(-30,-23){\psfig{file=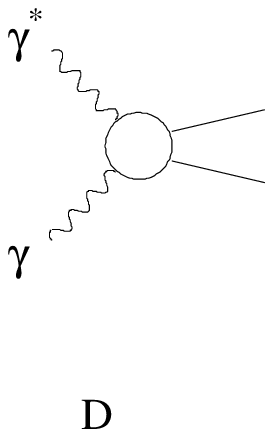,width=10.4cm}}
\put(10,-23){\psfig{file=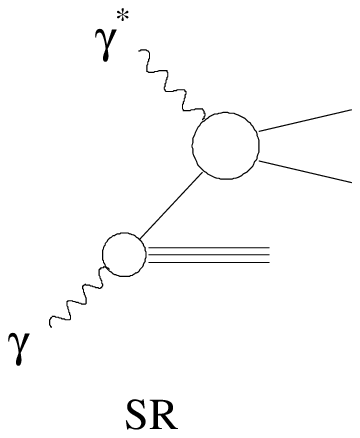,width=10.4cm}}
\put(50,-23){\psfig{file=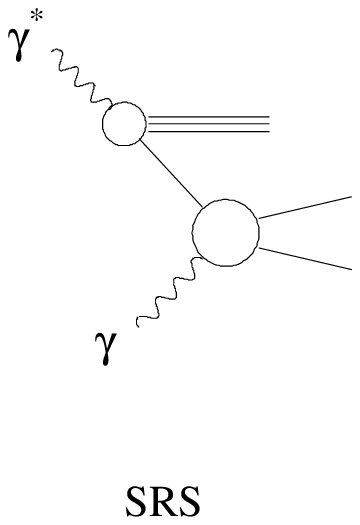,width=10.4cm}}
\put(90,-23){\psfig{file=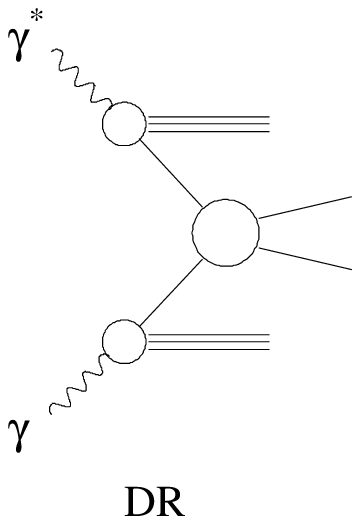,width=10.4cm}}
\end{picture}
	\caption{\label{sdr}\it The different components contributing
	to jet production in $\g^*\g$-scattering.}
\end{figure}

It is desirable to find alternative ways of probing the virtual photon
PDF's, either to test their universality or to further constrain their
$Q^2$-dependence. One possibility is the reaction
$\gamma^*(Q^2)+\gamma (P^2\simeq 0)\to \mbox{jets}(E_T)+X$, where the 
virtuality of the probing photon $Q^2$ has to be sufficiently small in
comparison with the transverse jet energy $E_T$ to allow for a
hadronic component in the virtual photon. The real photon with
virtuality $P^2\simeq 0$ involved in the collision has both a direct
pointlike and a resolved hadronic part. This reaction can be obtained
at $e^+e^-$ colliders by single-tag experiments, and preliminary
results have been reported from the OPAL collaboration at LEP
\cite{opal2} for $6\le Q^2\le 30$~GeV$^2$. For the large 
$Q^2$ region we have recently evaluated the NLO QCD corrections to
this process  and found them to be small and to improve the 
scale dependence \cite{12}. Moving towards the region of small virtualities,
especially for $Q^2\ll E_T^2$, one expects logarithms from the initial 
state $\gamma^*\to q\bar{q}$ splitting of the form $\ln (Q^2/s)$, where
$\sqrt{s}$ is the partonic center-of-mass energy, to become 
large and spoil the convergence of the perturbative expansion. A
procedure for subtracting these terms from the direct virtual photon
cross sections and absorbing them into the PDF of the resolved photon
has been worked out for the case of $eP$-scattering in \cite{2} in the
framework of the phase-space-slicing method in analogy to the
subtraction of $1/\e$ poles from photoproduction \cite{kk,kkk}. In
section 2 of this paper we will work out the subtraction procedure for the 
$\gamma^*\gamma$-scattering case, closely following the approach in
\cite{2}. With the subtraction performed, one has to
include a resolved virtual photon contribution into the jet
cross sections. In the deep-inelastic case as
described in \cite{12}, the virtual photon couples only directly and
the real photon can have a direct and a resolved component. These are
called the direct (D) and single-resolved (SR) contributions. In
addition to these, the so-called single-virtual resolved (SRS) and
double-resolved (DR) contributions occur, when the virtual resolved
photon components are included. The four contributions are shown in
Fig.~\ref{sdr}. In the SRS case, the real
photon interacts directly and the virtual photon is resolved. For this
case the  NLO matrix elements can be taken from
photoproduction \cite{kk,kkk,15}. Finally, in the DR case both photons
are resolved, for which the  matrix elements are described in
\cite{kkk}. Since NLO calculations for the case of
$\gamma\gamma$-scattering with both  photons being on-shell, i.e., 
$Q^2\simeq P^2\simeq 0$, are available in the literature
\cite{kkk,15,14}, we compare our results in the limiting case
$Q^2\to 0$ with these calculations. 

It has been shown in \cite{10,10b,11} that the parton content of virtual
photons is not solely described by purely perturbative contributions
in the region $\Lambda_{QCD}^2\ll Q^2 \ll \mu^2$, where $\mu^2$ is the
probing scale, in contrast to the expectations from \cite{p2}. However, 
as has been pointed out in \cite{10b,13}, it is not clear for which values
of $Q^2$ (and $\mu^2$) the non-perturbative part is relevant and down
to which value of $Q^2$ one should trust perturbation theory. In
section 3 we will therefore numerically study the perturbative
stability of the $\gamma^*\gamma\to\mbox{jets}$ cross sections in the
region $\Lambda_{QCD}^2\ll Q^2 \simlt E_T^2$, where $E_T$ is the
typical hard scale in jet production. We study the $K$ factors for
different scales and virtualities and compare the scale dependences of
the unsubtracted jet cross sections D and SR with those obtained after the
subtraction of the logarithms and inclusion of resolved components SRS
and DR. From this comparison we can deduce the regions of photon
virtuality where fixed order perturbation theory gives reliable
predictions for jet cross sections and where, on the other hand, the
parton content of a virtual photon is relevant.

Finally, in section 4, we make predictions for low $Q^2$ jet
production including a virtual resolved photon component as functions
of $Q^2$ and $E_T$ and discuss whether the resolved component can be
observed at LEP energies. Further numerical results can be found in
\cite{12a}. The paper ends with a summary of the results and conclusions.

\clearpage

\section{Low $Q^2$ jet cross sections}

Taking over the conventions given in \cite{12}, the process we are
interested in can be written as 
\equ{ee}{ e^+(k_a) + e^-(k_b) \longrightarrow e^+(k_a') + e^-(k_b') +
  \mbox{jets}(E_T) + \mbox{X}  \quad ,}
with the subprocess $\g^*(q_a)+\g (q_b) \to \mbox{jets} +
\mbox{X}$, where $q_a=k_a-k_a'$ and $q_b=k_b-k_b'$. The virtualities
are given by $Q^2 = -q_a^2$ and $P^2 = -q_b^2=0$. The
cross section for the process (\ref{ee}) at large $Q^2$ is given by
the convolution 
\begin{eqnarray}
  \frac{d\sigma_{e^+e^-}}{dQ^2dy_ady_b} = \sum_b \int dx_b
  F_{\gamma /e^-}(y_b) f_{b/\gamma}(x_b)
  \frac{\alpha}{2\pi Q^2} \left[ \frac{1+(1-y_a)^2}{y_a}
  d\sigma^U_{\gamma^*b} \right. \nonumber \\ 
  + \left. \frac{2(1-y_a)}{y_a} d\sigma^L_{\gamma^*b} \right] \ ,
\label{e+e-}
\end{eqnarray}
where $y_a = (q_ak_b)/(k_ak_b)$. The variable $y_b\in [0,1]$ describes
the momentum fraction of the real photon in the electron. The
momentum fraction of a parton in the real photon is $x_a\in [0,1]$ and
the PDF of the real photon is $f_{b/\gamma}(x_b)$. The indices $U$ and $L$
denote the unpolarized and longitudinally polarized virtual photon
contributions. Finally, the function $F_{\g /e^-}(y_b)$ is the
Weizs\"acker-Williams approximation for the real photon \cite{wwill}, 
\begin{equation} \label{ww}
 F_{\g /e^-}(y_b) = \frac{\al}{2\pi} \frac{1+(1-y_b)^2}{y_b}
  \ln \left( \frac{E_e^2\theta^2_{max}}{m_e^2} \right) \ , 
\end{equation}
where $m_e$ is the electron mass and $\theta_{max}$ is the maximum
scattering angle of the untagged electron. An improved version of the
equivalent-photon approximation in $e^+e^-$ collisions has been
derived in \cite{schuler}, leading however only to small
corrections. We use the simpler formula (\ref{ww}), since our studies
are exploratory and we do not compare with data.

The definition of the partonic cross sections $d\si^{U,L}_{\gamma^*b}$ are 
given in \cite{12}. For the SR case, the NLO calculations can be taken
from \cite{graud}, whereas the NLO calculations for the D case have been
presented in \cite{12}; both calculations employ the
phase-space-slicing method to extract the singular phase-space regions
of the real corrections. The singular integrals of both the real and
the virtual corrections are handled in dimensional
regularization. Most of the $1/\e^n$ poles for the real and virtual
corrections cancel and the remaining initial state corrections are
subtracted into the real photon PDF. We are in the following interested
in the initial state splitting for the virtual photon, $\gamma^*\to
q\bar{q}$, which has been evaluated in \cite{2} for the SR
contributions. Here, we repeat this calculation for the D 
process\footnote{Note, that in \cite{2} the photon virtuality
corresponding to our $Q^2$ was called $P^2$.}
\equ{II}{ \gamma^*(q_a)+\gamma(q_b) \to q(p_1)+\bar{q}(p_2)+g(p_3) \ .}
The first step is to extract from the $2\to 3$ matrix elements of the
reaction (\ref{II}) a term with the characteristic denominator from
the $\gamma^*\to q\bar{q}$ splitting which gives the singular
contribution in the limit $Q^2\to 0$. We call this term $H_K$. In
the same singular limit the three-body  phase space of the $q\bar{q}g$
final state, $d\mbox{PS}^{(3)}$, factorizes according to
$d\mbox{PS}^{(3)} = d\mbox{PS}^{(2)} d\mbox{PS}^{(r)}$, with
$d\mbox{PS}^{(2)}$ being the usual two-body phase-space and
$d\mbox{PS}^{(r)}$ being the phase-space of the the singular region
\cite{2,kkk,graud}. We define the variable $z_a \equiv
(p_2p_3)/(q_aq_b) \in [\eta_a,1]$ that gives the fraction 
of the momentum $q_a$ that participates in the subprocess after a
particle has been radiated in the initial state. The variable
$\eta_a\in [0,1]$ is connected to $z_a$ through $\eta_a=x_az_a$. The
term $H_K$ is integrated over the singular phase space up to a cut-off
$y_s$ with the result
\equ{I}{ \int d\mbox{PS}^{(r)} H_K = \frac{\al_s}{2\pi}(4\pi\al
 Q_i^2)^2  \int\limits_{\eta_a}^1 \frac{dz_a}{z_a} \ 
	2N_CC_FM(Q^2)\ T_\g (s,t,u)  \quad , } 
where
\equ{m(P2)}{ M(Q^2) = \frac{1}{2N_C} P_{q\leftarrow\g}(z_a) \ln\left(
  1 + \frac{y_ss}{z_aQ^2} \right)  }
is the singular term and
\equ{}{ T_\g (s,t,u)=-\frac{s}{u}-\frac{u}{s} }
is the LO photon-parton scattering matrix element where $s,t$ and $u$
are the usual Mandelstam variables. The photon splitting function is
given by $P_{q\leftarrow \g}(z_a) = N_C \left( z_a^2 + (1-z_a)^2 \right)$.
The term (\ref{m(P2)}) is large for $Q^2\ll s$ and singular for $Q^2=0$, as
expected. It is the same universal term as obtained in \cite{2} for 
the $eP$-case. We therefore introduce, in accordance with \cite{2}, the
subtraction term 
\equ{st}{ \Gamma_{q\leftarrow\g}(z_a,M_\g^2) = \ln \left( 
	\frac{M_\g^2}{Q^2(1-z_a)}\right) P_{q\leftarrow\g}(z_a) -N_C }
which is to be absorbed into the PDF of the virtual photon. After this
subtraction the remaining finite term in $M(Q^2)$ yields
\equ{}{ M(Q^2)_{\overline{MS}} = -\frac{1}{2N_C}
  P_{q\leftarrow\g}(z_a)\ln\left(\frac{M_\g^2z_a}{(z_aQ^2+y_ss)(1-z_a)} 
\right)  + \frac{1}{2} \ . }
In addition to the singular term $\ln (M_\g^2/Q^2)$ two finite terms
have been subtracted in order to achieve the $\overline{\mbox{MS}}$
factorization for $Q^2\ne 0$ \cite{2}. It is defined by the
requirement that the remaining finite term $M(Q^2)_{\overline{MS}}$ 
is equal to the finite term obtained after factorization of the real
photon initial state singularities, which can be found in
\cite{kk,kkk,15}. Of course, this has consequences concerning the
selection of the PDF of the virtual photon, the details of which can
be found in \cite{2}. This completes the calculation of the
contribution from the virtual photon initial state singularity. Note
that the above described subtraction only concerns the transversely
polarized virtual photons in reaction (\ref{II}), since  the
contributions for longitudinal photons vanish in the limit $Q^2\to
0$. Furthermore, up to now no virtual photon structure function for
longitudinal photons has been constructed. The final formula for jet cross
sections including resolved virtual photon contributions in $e^+e^-$
scattering at low $Q^2$ is a generalization of (\ref{ee}) and reads
\begin{eqnarray} 
  \frac{d\sigma_{e^+e^-}}{dQ^2dy_ady_b} = \sum_{a,b} \int dx_adx_b
  F_{\gamma /e^-}(y_b) f_{b/\gamma}(x_b)
  \frac{\alpha}{2\pi Q^2} \left[ \frac{1+(1-y_a)^2}{y_a}
  f^{vir}_{a/\gamma^*}(x_a,Q^2)d\sigma_{ab} \right. \nonumber \\ 
  + \left. \frac{2(1-y_a)}{y_a} \delta (1-x_a)d\sigma^L_{\gamma^*b}
  \right] \ , \label{ee2}
\end{eqnarray}
where $f^{vir}_{a/\gamma^*}(x_a,Q^2)$ is the virtual photon PDF 
and $x_a\in [0,1]$ describes the momentum fraction of the parton
in the virtual photon. The direct photon interactions are included in
this formula through delta functions. For the direct virtual photon one has  
the relation $f^{vir}_{\gamma^*/\gamma^*}d\sigma_{ab} = \delta
(1-x_a)d\sigma^U_{\gamma^*b}$, whereas for the direct real photon
the relation is $f_{\gamma/\gamma}d\sigma_{ab} = \delta
(1-x_b)d\sigma_{\gamma b}$, where $d\sigma_{ab}$ refers to the
partonic cross section. 

Formula (\ref{ee2}) is implemented in the fixed higher order program
{\tt JetViP} \cite{jv}, with which all numerical results in this
paper are produced. The input parameters for all numerical studies are
the following. We assume LEP2 conditions, i.e., the energies of the
incoming leptons are $E_a=E_b=91.5$~GeV. We integrate over the full
range of $y_a,y_b\in [0,1]$ and use the value $\theta_{max}=0.025$ in
(\ref{ww}). For the real photon we always employ the PDF's of Gl\"uck,
Reya and Vogt (GRV) \cite{grv}, whereas for the virtual photon we will
use the new GRS \cite{10b} parametrization and the SaS1D PDF \cite{11} 
transformed to the $\overline{\mbox{MS}}$-scheme \cite{2}. The GRS
PDF's are constructed for $N_F=3$ flavours, the production of the heavier
$c$ and $b$ quarks is supposed to be added as predicted by fixed order
perturbation theory. We calculate
our cross sections with $N_F=5$ flavours and use the two-loop
formula for the strong coupling constant without threshold effects
with $\Lambda_{QCD}^{(5)}=153$~MeV, even for the LO results. The cross
sections are plotted in the $\gamma^*\gamma$ cms, where also jets are
defined. We use the Snowmass accord \cite{snow}, where two partons $i$
and $j$ are recombined, if for both partons the condition $R_{i,j}<R$
is fulfilled, where $R_{i,J}=\sqrt{(\eta_i-\eta_J)^2+(\phi_i-\phi_J)^2}$ 
and  $\eta_J,\phi_J$ are the rapidity and the azimuthal angle of the
combined jet respectively, defined as
\begin{eqnarray}
 E_{T_J} &=& E_{T_1} + E_{T_2} \\
 E_{T_J}\eta_J  &=& E_{T_1}\eta_1 + E_{T_2}\eta_2 \quad , \\
 E_{T_J}\phi_J  &=& E_{T_1}\phi_1 + E_{T_2}\phi_2 \quad .
\end{eqnarray}
We choose $R=1$.

\begin{figure}
  \unitlength1mm
  \begin{picture}(122,65)
    \put(-6,-65){\epsfig{file=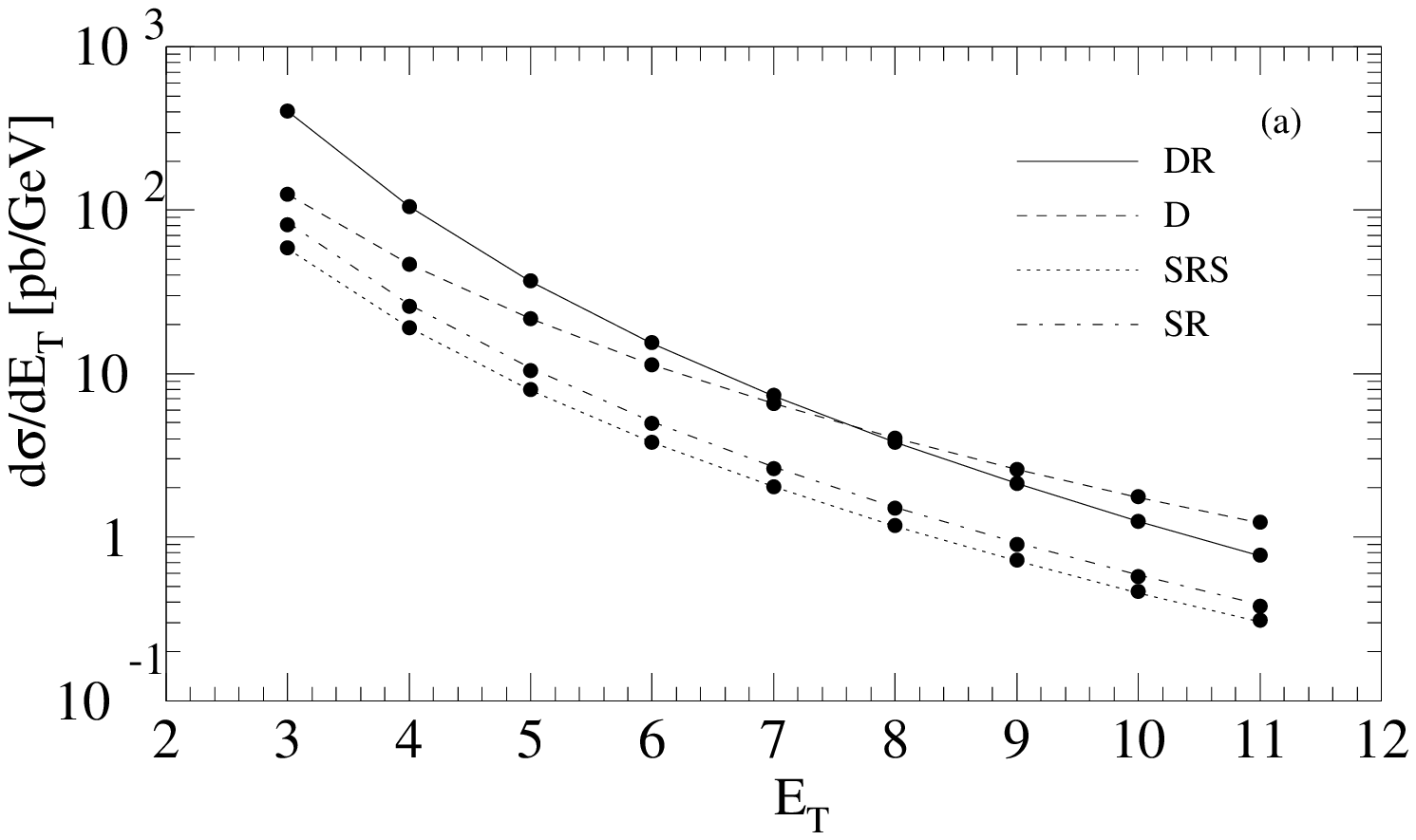,width=9.5cm,height=14cm}}
    \put(79,-65){\epsfig{file=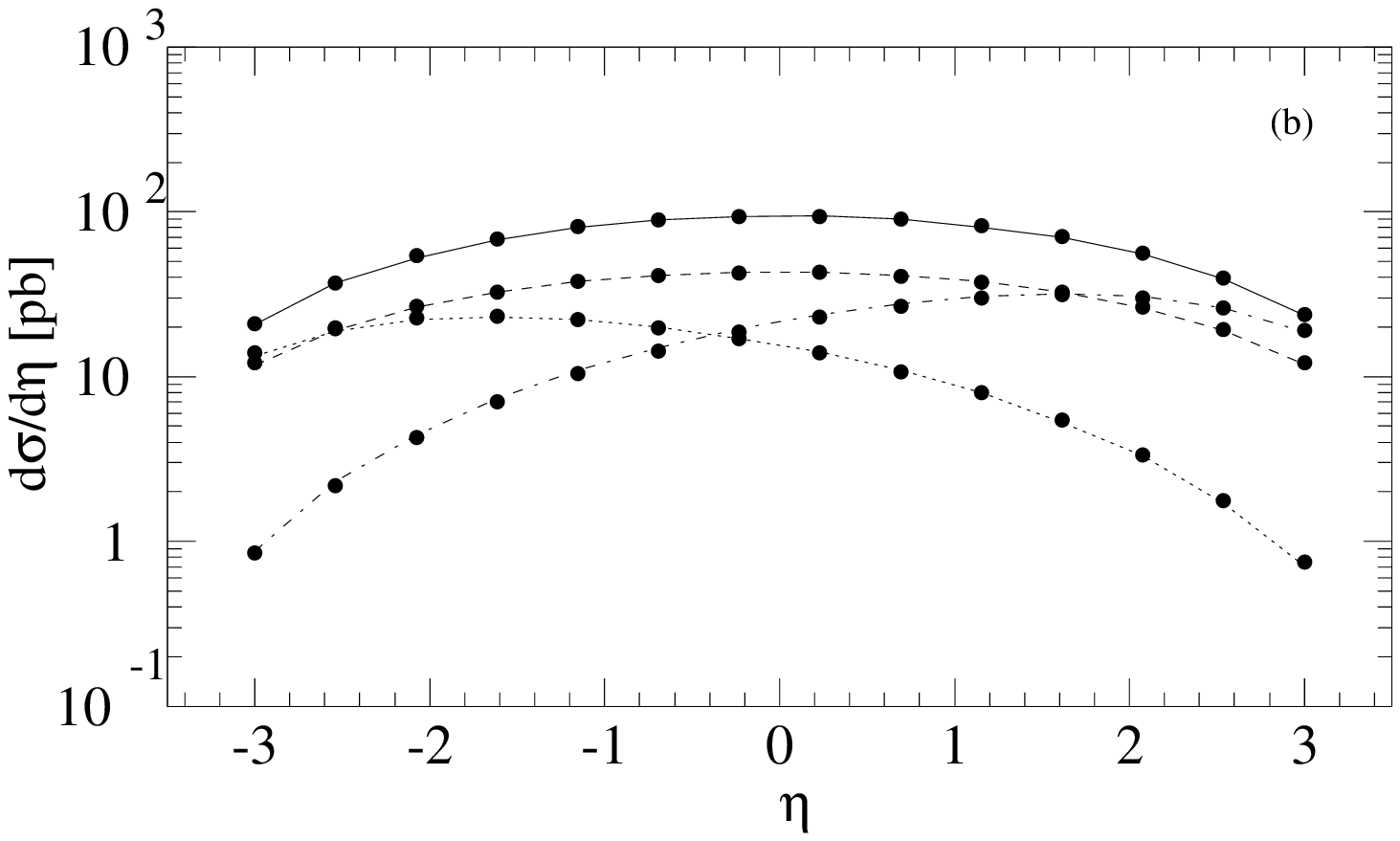,width=9.5cm,height=14cm}}
  \end{picture}
  \caption{\label{f0}\it Comparison of single inclusive jet cross sections
  in the limit $Q^2\to 0$ (lines) with the calculations of Kleinwort
  and Kramer (dots). The full lines give the DR, the dashed the D, the
  dotted the SRS and the dash-dotted the SR contributions. (a) $E_T$
  distribution with $|\eta|<2$; (b) $\eta$ distribution with
  $E_T>3$~GeV.}
\end{figure}

As a numerical check of our calculations and of the consistency of the
$\overline{\mbox{MS}}$-scheme for virtual photons, we compare in
Fig.~\ref{f0}~a,b the cross section (\ref{ee2}) in the limit $Q^2\to 0$ 
with existing calculations for photon-photon scattering \cite{15}. 
On the virtual photon side, the $Q^2$ is integrated out by using the
Weizs\"acker-Williams approximation \cite{wwill} with $Q^2_{max}=1$
GeV$^2$. In Fig.~\ref{f0}~a the comparison is made for $d\si/dE_T$ as a
function of $E_T$, where the rapidity has been integrated out in the
range $\eta \in [-2,2]$. The dots are the results from Kleinwort and
Kramer \cite{15}, whereas the curves are the predictions from {\tt JetViP}, 
for all four components, D, SR, SRS and DR, as discussed above, now
using the SaS PDF's. One sees a perfect agreement. The SR and SRS
distributions are very close to each other, but do not coincide
exactly, since different PDF's are employed for the real and the
virtual photon. However, the  SRS and SR cross sections
agree exactly on the partonic level. The excellent
agreement seen in the $E_T$ distributions holds also for the rapidity
distribution $d\si/d\eta$ of Fig.~\ref{f0}~b for all four
components. The transverse energy in these curves has been integrated
out with $E_T>3$ GeV. The SR and SRS components have their respective
maxima at opposite sides of the rapidity range. The SR distribution is
peaked at positive $\eta$'s, since the direction of the virtual photon
was taken to be the positive $z$-axis. In photoproduction, the DR
component gives considerable contributions in the low $E_T$ region, as
can be seen from the $E_T$ spectra. The D cross section only starts to
dominate for $E_T$ larger than $10$~GeV. The SR and SRS components are
of minor importance.

\section{Direct vs.\ resolved virtual photon approaches} 

We now compare the two approaches for calculating jet cross sections
at small $Q^2$, namely considering the direct coupling of the virtual
photon only and, on the other hand, including the resolved virtual
photon components. We cover the region $\Lambda_{QCD}^2\ll Q^2 \simlt E_T^2$ 
where one could expect contributions from resolved virtual photons to
be of importance. For the discussion in this section we will only use
one of the virtual photon PDF's, namely those of SaS,
which are built for $N_F=5$ flavours. The input parameters for the
following plots are as described in the previous section. 

\begin{figure}
  \unitlength1mm
  \begin{picture}(122,65)
    \put(30,-65){\epsfig{file=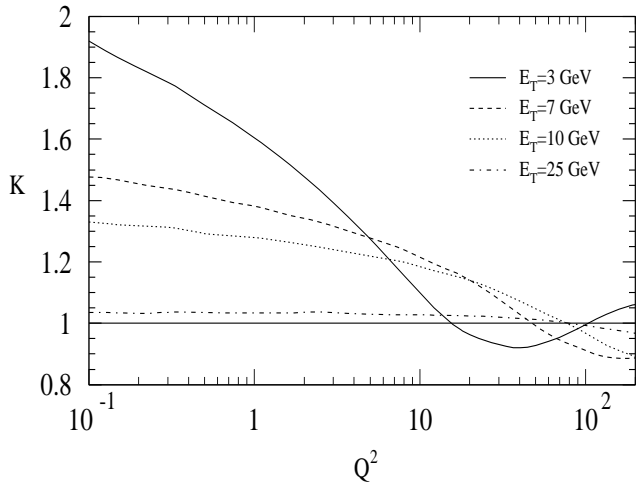,width=9.5cm,height=14cm}}
  \end{picture}
  \caption{\label{f1}\it Ratio of NLO to LO single jet cross sections
	integrated over $|\eta|<2$ for $E_T=3,7,10$ and $25$~GeV as a
	function of $Q^2$.}
\end{figure}

In Fig.~\ref{f1} we have plotted the ratio of the single inclusive jet
cross sections
\begin{equation}
	K = \left(\frac{d\si^{NLO}}{dE_T}\right) \bigg/ 
	\left(\frac{d\si^{LO}}{dE_T}\right) \ , 
\end{equation}
where $d\si^{NLO}$ contains all Born terms. The cross sections are 
obtained by integrating over the rapidity-range $|\eta|<2$ and
for four different values of transverse energy, $E_T=3,7,10$ and
$25$~GeV, in the region  $Q^2\in [0.1,200]$~GeV$^2$, where the lower
limit is roughly  $Q^2_{min}\simeq10\times\Lambda_{QCD}^2$. The
renormalization and factorization scales $\mu_R$ and $\mu_F$ are
set equal to $E_T$. We observe that $K$ is close to unity for
all four curves for $Q^2\simgt 30$~GeV$^2$ which indicates the perturbative
stability of the cross sections in the deep-inelastic region. The 
ratio exceeds $1$ below $Q^2=E_T^2$ and rises monotonically towards
smaller $Q^2$ for the three smaller $E_T$'s. For $E_T=25$~GeV the NLO
terms give only very small corrections below 5\% in the whole $Q^2$
region down to $Q^2_{min}$.  
For $E_T=3$~GeV the rise towards smaller $Q^2$ is strongest and leads
to a NLO correction of nearly $100$~\% near $Q^2\simeq Q^2_{min}$, but
is already around $50$~\% for $Q^2\simeq E_T^2/5$. For the larger
$E_T$ values the NLO corrections do not rise that dramatically,
however below $Q^2=10$~GeV$^2$ the corrections exceed $20$~\% for
$E_T=7$ and $10$~GeV.

\begin{figure}[hhh]
  \unitlength1mm
  \begin{picture}(122,125)
    \put(-4,-5){\epsfig{file=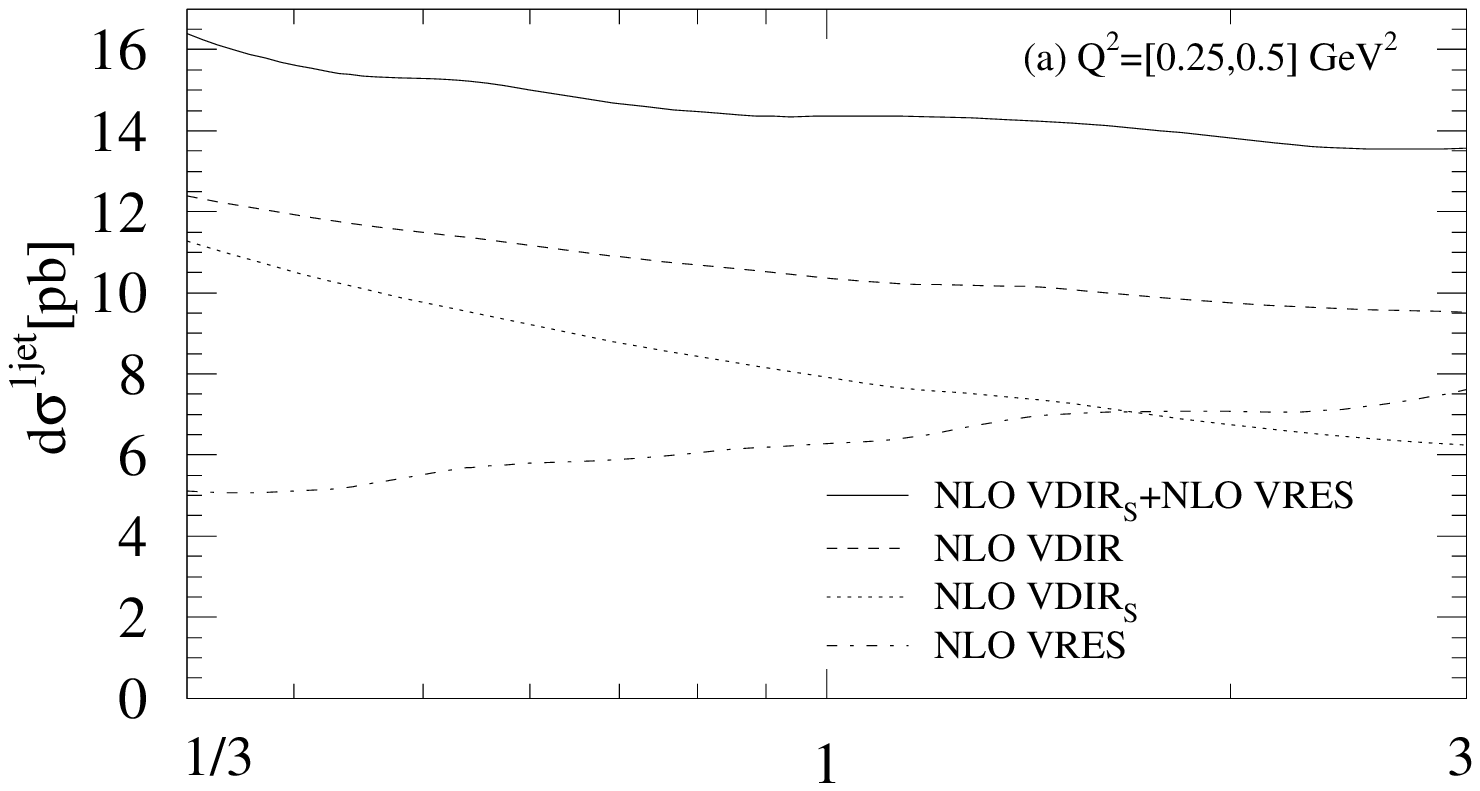,width=9.5cm,height=14cm}}
    \put(78,-5){\epsfig{file=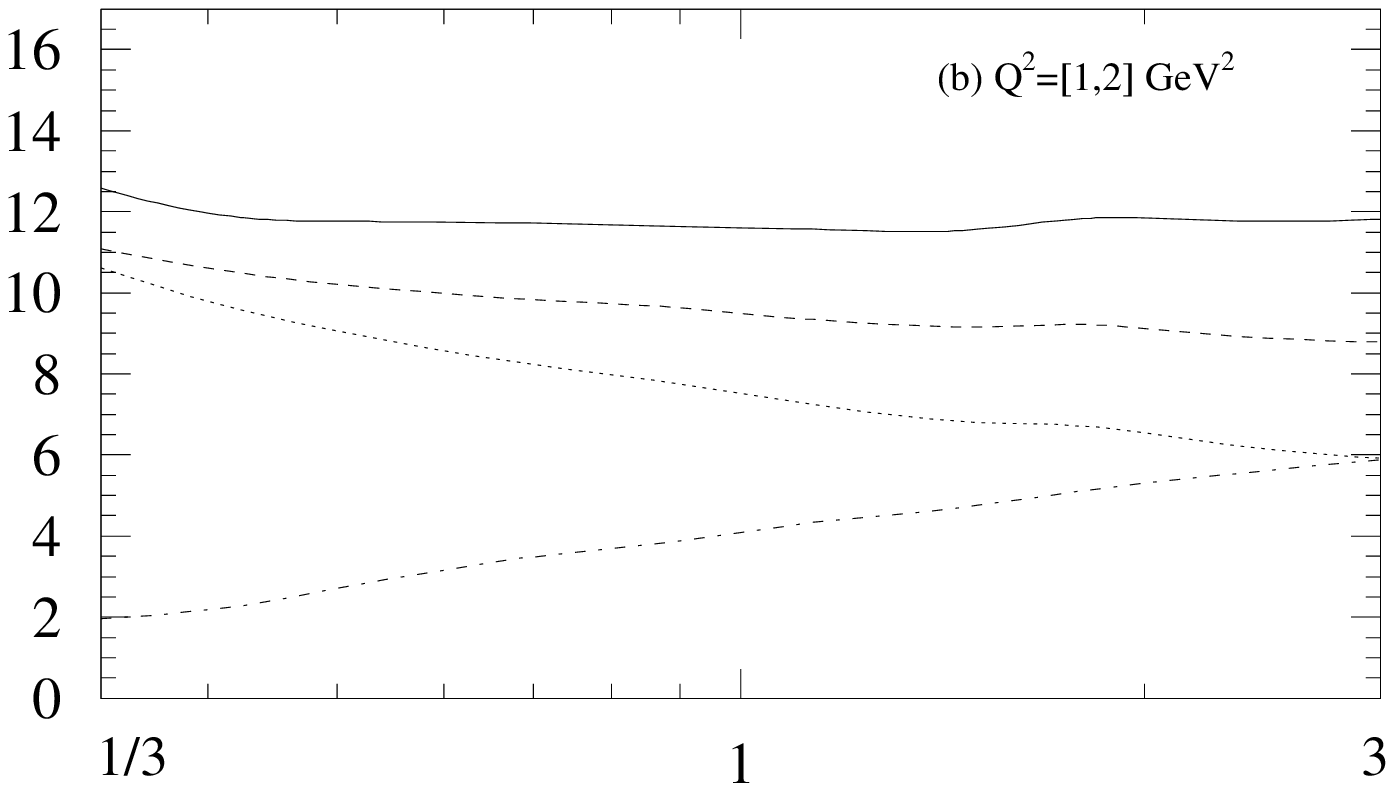,width=9.5cm,height=14cm}}
    \put(-4,-65){\epsfig{file=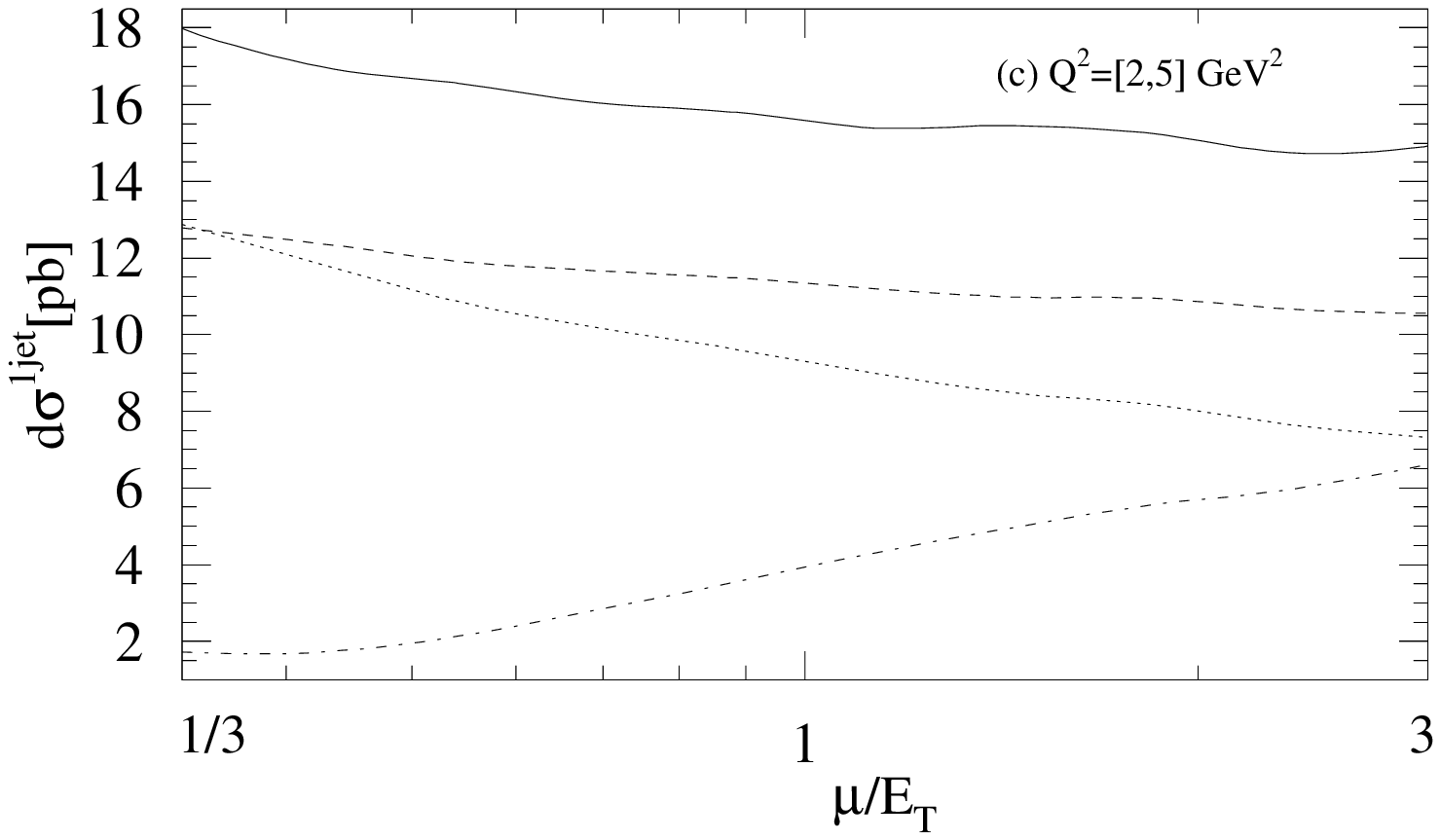,width=9.5cm,height=14cm}}
    \put(78,-65){\epsfig{file=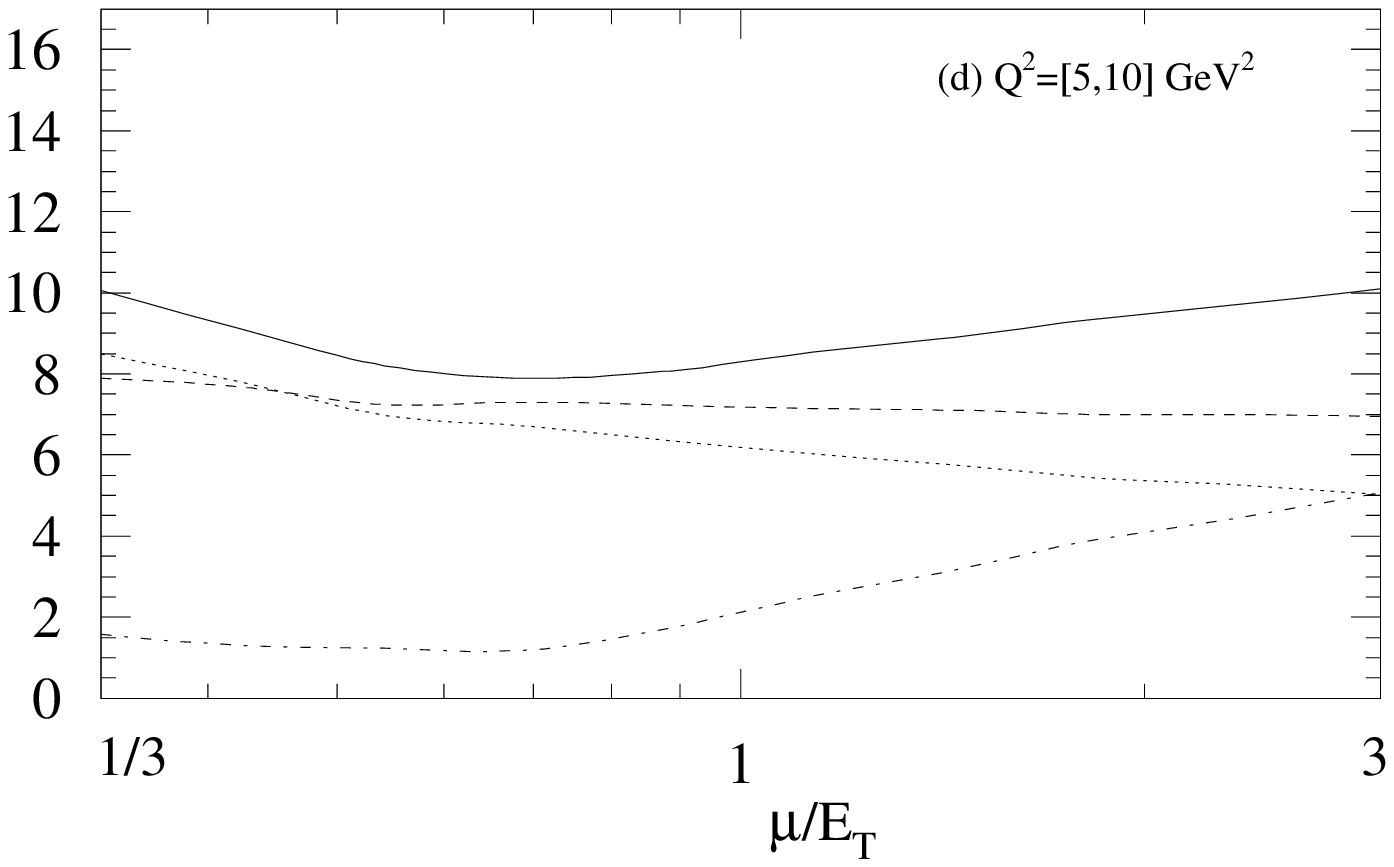,width=9.5cm,height=14cm}}
  \end{picture}
  \caption{\label{f2}\it Scale dependences of single inclusive jet cross
  sections integrated over $|\eta|<2$ and $E_T>3$~GeV as a
  function of $\mu/E_T$ for (a) $Q^2\in [\frac14,\frac12]$~GeV$^2$,
  (b) $Q^2\in [1,2]$~GeV$^2$, (c) $Q^2\in [2,5]$~GeV$^2$ and (d)
  $Q^2\in [5,10]$~GeV$^2$. The dashed line is the NLO VDIR which has
  to be compared with the full line, giving the sum of the subtracted
  virtual direct (NLO VDIR$_S$, dotted line) and the NLO virtual
  resolved (NLO VRES, dash-dotted line).} 
\end{figure}

We take out the case $E_T=3$~GeV to make a more detailed study of
the perturbative stability by looking at the scale dependence of the
cross sections. In Fig.~\ref{f2} a--d we show the single
jet inclusive cross section integrated over $|\eta|<2$ and
$E_T>3$~GeV as a function of $\mu/E_T\in [\frac13,3]$, where
$\mu=\mu_R=\mu_F$, i.e., renormalization and factorization scales are
varied at the same time. The curves are shown in the four bins of virtuality
$Q^2\in [0.25,0.5], [1,2], [2,5]$ and $[5,10]$~GeV$^2$. Since we are
interested only in comparing direct and resolved virtual photon
contributions, we have introduced the following combinations of the
four components D, SR, SRS and DR: the {\it virtual direct} (VDIR) is
the sum of D and SR components, whereas the {\it virtual resolved}
(VRES) is the sum of SRS and DR components. The component labeled
VDIR$_S$ is the virtual direct component after subtraction of the term
(\ref{st}) in the D and the corresponding term in the SR component, to
be found in \cite{2}. 

The scale variation of the NLO VDIR contribution between the smallest
and largest $\mu$ value in Fig.~\ref{f2}~a amounts 30\% and goes down
to around 15\% in the largest $Q^2$ bin. We have checked that the
renormalization scale dependence alone gives a variation of 60\% for the
smallest and 25\% for the largest $Q^2$ bin, i.e., the scale variation
behaviour becomes stable and well behaved for $Q^2$ approaching
$E_T^2$. Most of the variation with the renormalization scale stems from
the SR component, whereas the D component is much more stable and
varies at most 10\%. This is understandable, since the D component is
only ${\cal O}(\al_s)$ in the strong coupling, whereas the SR
component is ${\cal O}(\al_s^2)$. The rather strong variation of the
VDIR component at small $Q^2$ values together with the large $K$
factor observed above, indicates the need for a resummed approach at
small $Q^2$. We have therefore also plotted the NLO VRES component in
Fig.~\ref{f2} a--d. To avoid double counting, the VRES component can
only be added after the perturbative $\gamma^*\to q\bar{q}$ splitting
has been subtracted from VDIR (giving VDIR$_S$) since these terms are
contained in the pointlike part of the 
virtual photon PDF. One sees two effects of including the VRES
component. First, the scale variation is considerably reduced in the
first two $Q^2$ bins, namely by a factor of $2$ in the first bin and
even a factor of 4 in the second bin. For the third bin the scale
variation of the $\mbox{SUM}=\mbox{VDIR}_S+\mbox{VRES}$ is more or
less the same as for the VDIR component alone, i.e., around 20\%. For
$Q^2\in [5,10]$~GeV$^2$ the SUM prediction becomes unstable for
$\mu\simlt 0.7E_T$, since in this region $Q^2>\mu^2$ and the
contribution from the virtual photon PDF is very small. But even for
$\mu >E_T$ the scale variation of SUM is much larger than for the VDIR
alone. The second observation is that including the VRES component in
NLO gives a relatively large correction to the pure VDIR component
of 20--30\%, up to 40\% in bin c.

\begin{figure}[hhh]
  \unitlength1mm
  \begin{picture}(122,125)
    \put(-4,-5){\epsfig{file=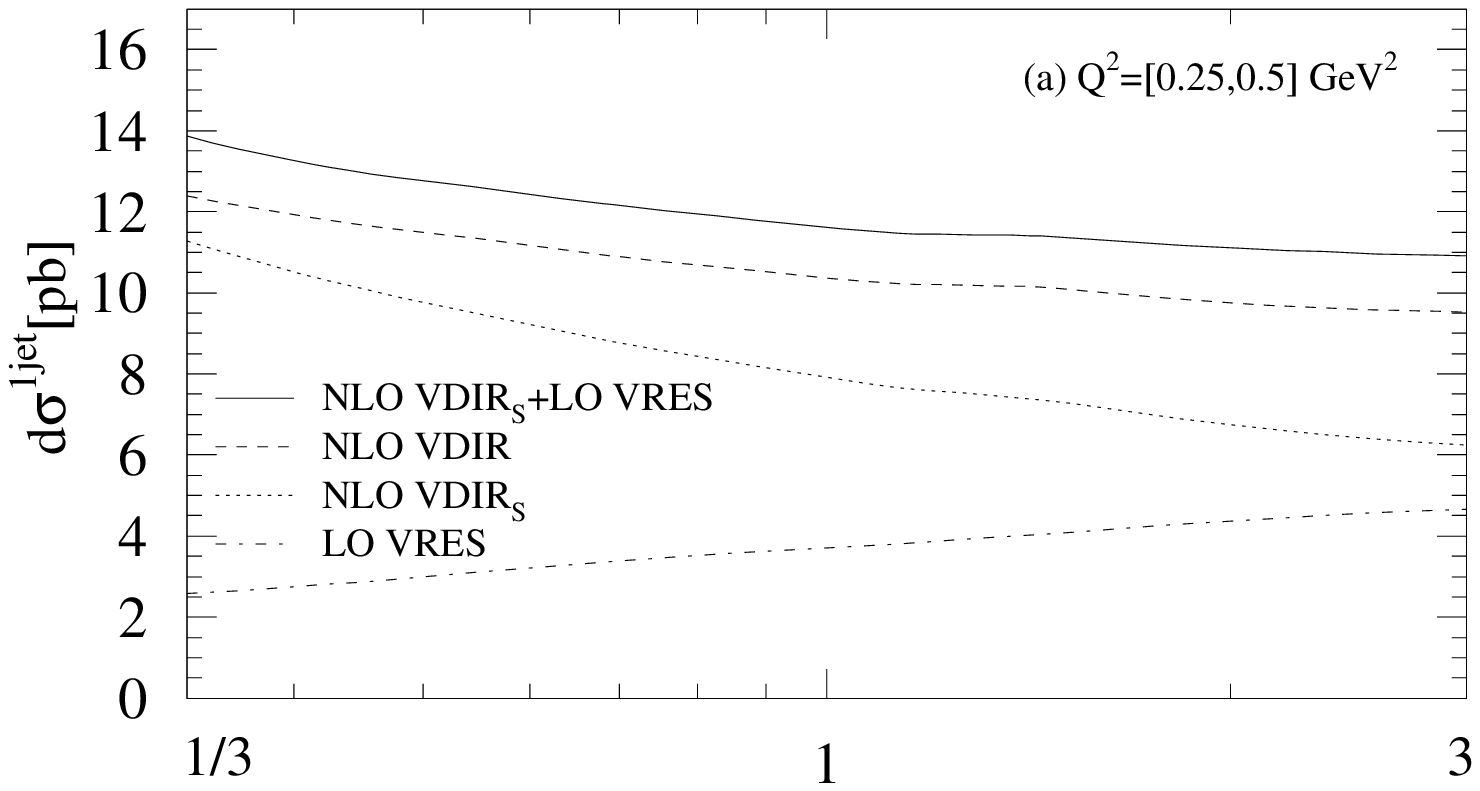,width=9.5cm,height=14cm}}
    \put(78,-5){\epsfig{file=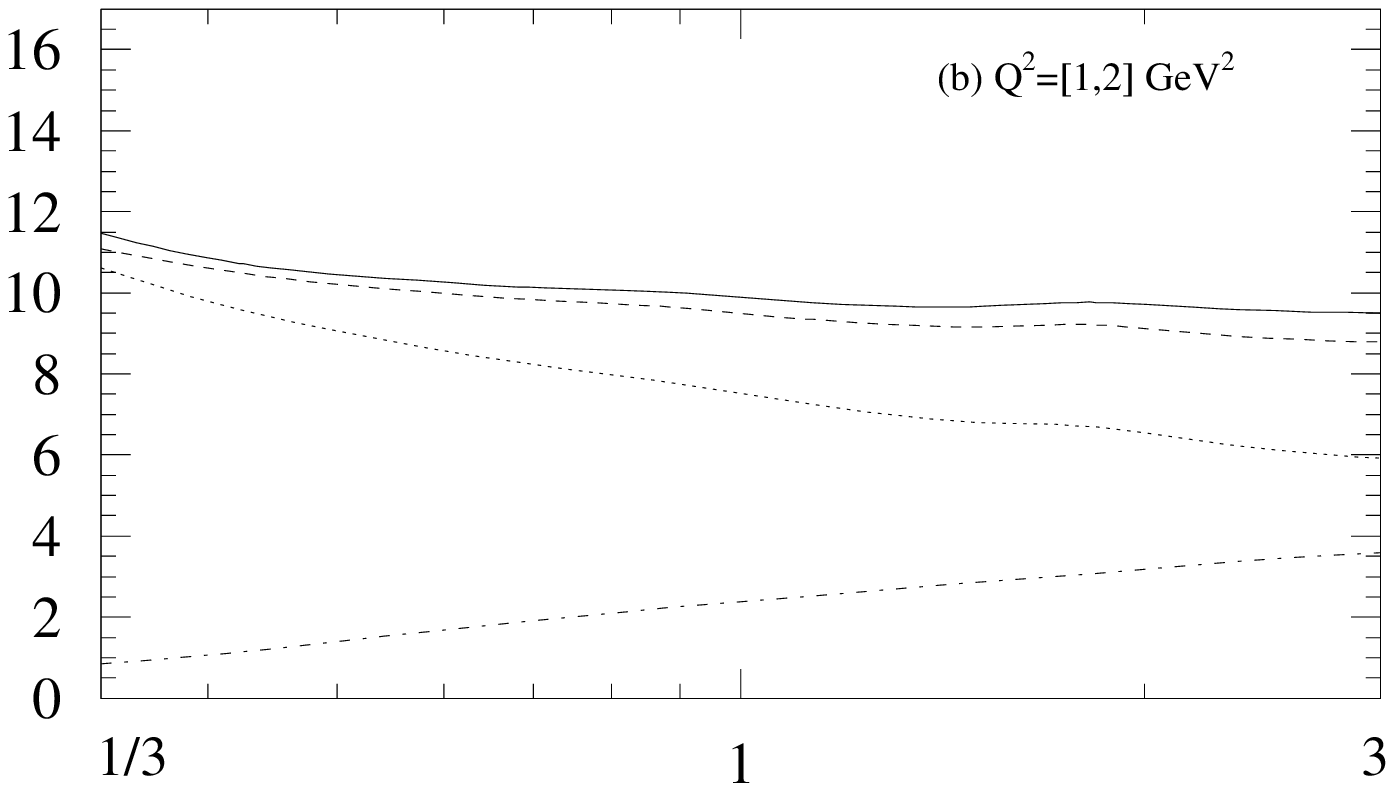,width=9.5cm,height=14cm}}
    \put(-4,-65){\epsfig{file=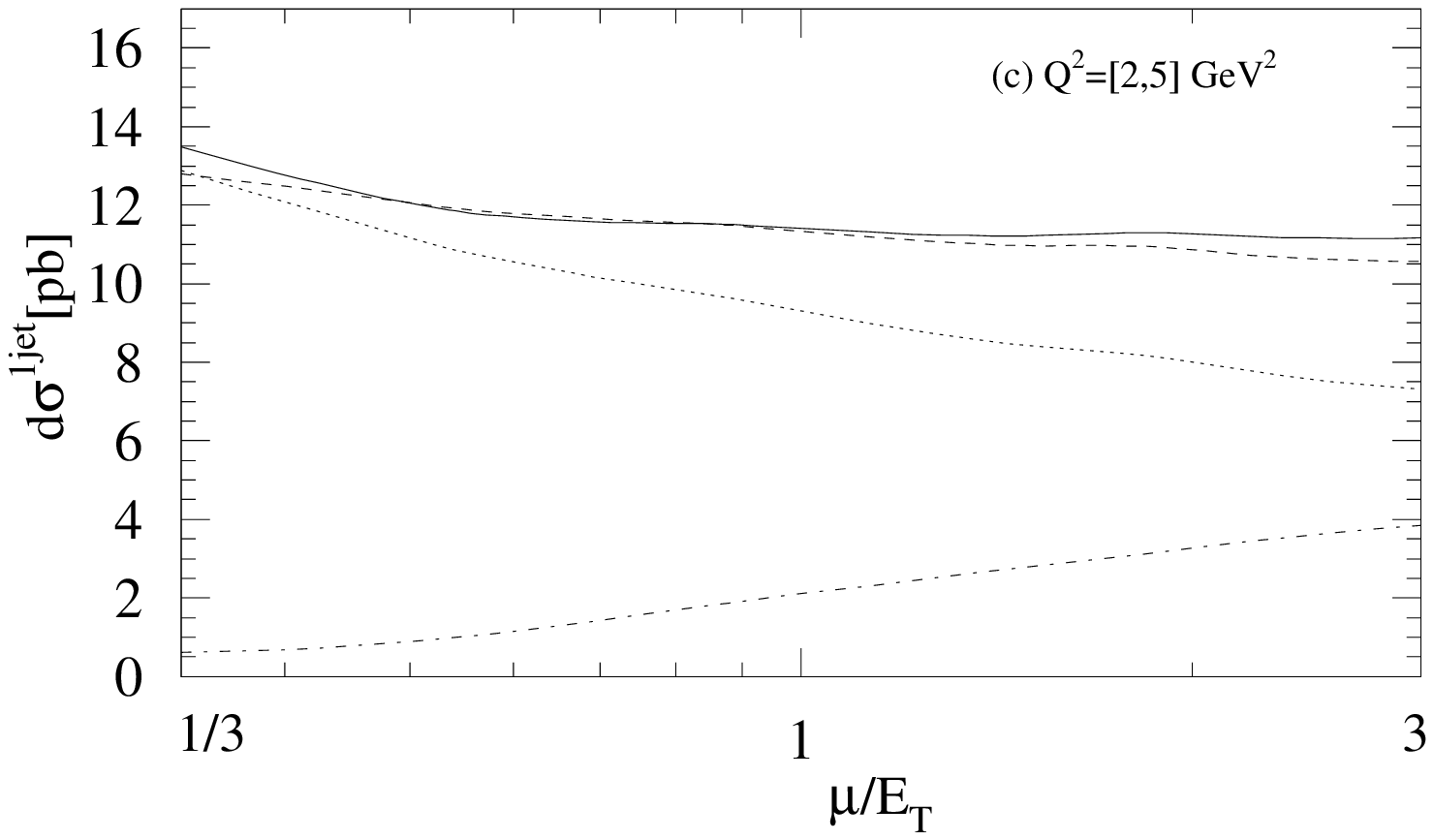,width=9.5cm,height=14cm}}
    \put(78,-65){\epsfig{file=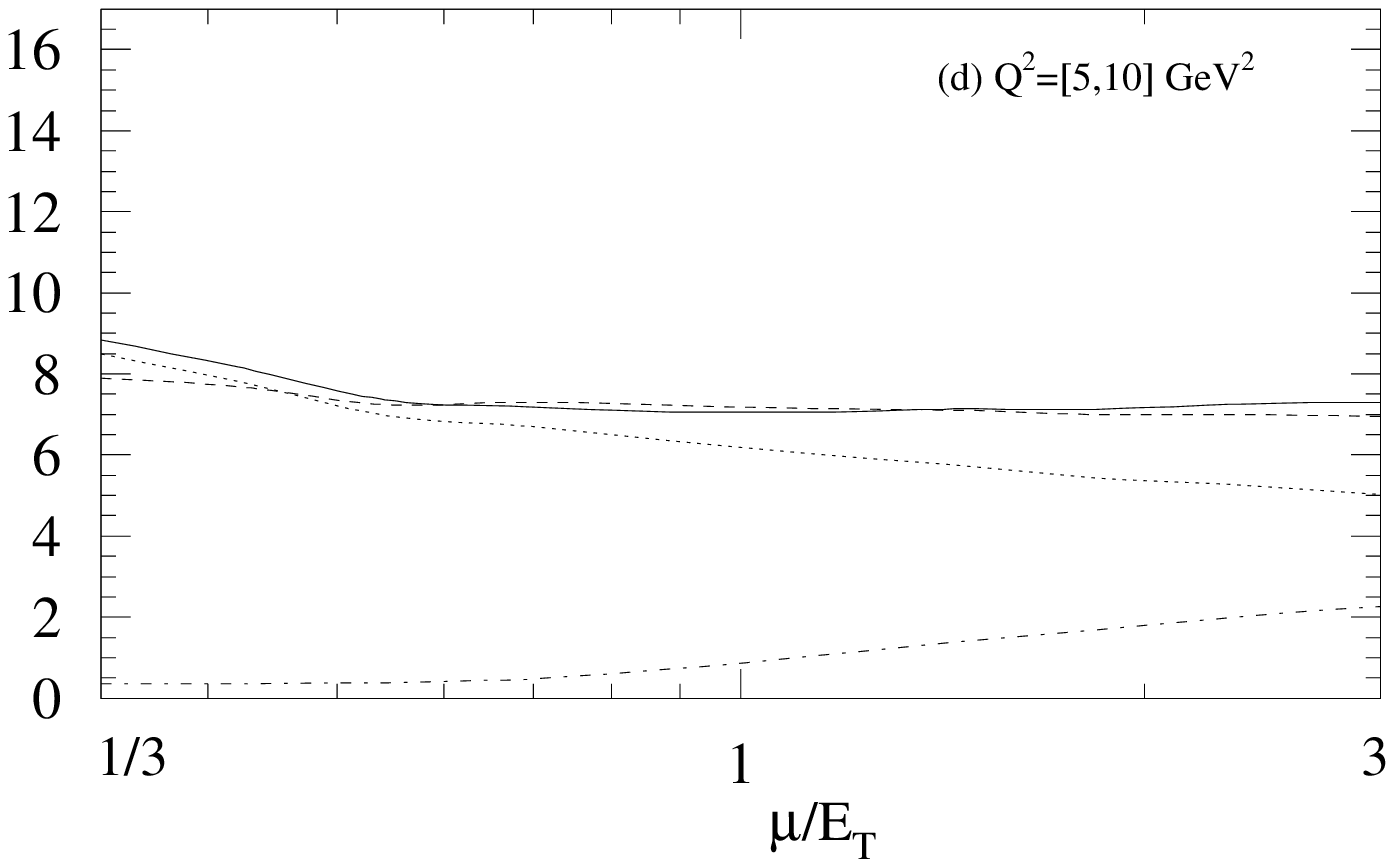,width=9.5cm,height=14cm}}
  \end{picture}
  \caption{\label{f3}\it Scale dependences of single inclusive jet cross
  sections with the same conditions as in Fig.~\ref{f2} a--d, only
  here the virtual resolved contribution is included in LO instead of NLO.}
\end{figure}

To disentangle the effects of resummation and non-perturbative parts
in the virtual photon PDF from those of higher order contributions in
the matrix elements we plot in Fig.~\ref{f3} a--d the same curves as in
Fig.~\ref{f2} a--d with the VRES component in LO only. The sum of NLO
VDIR$_S$ and LO VRES gives rather small corrections to the pure VDIR
of about 15\% in the smallest $Q^2$-bin and only 5\% in bin b. In the
bins c and d the two approaches, NLO VDIR and NLO VDIR$_S$+LO VRES,
give nearly the same results. The conclusion from this is that the
virtual photon PDF is to a very large extent given by the splitting
term for $Q^2>2$~GeV$^2$ at $E_T\simeq 3$~GeV, i.e., for
$Q^2\simgt\frac15\mu^2$. At smaller virtualities one sees effects from
resummation and some non-perturbative input. Only small improvements
can be seen from including the VRES component in the bins a and b with
respect to the scale variation, in contrast to the findings of 
Fig.~\ref{f2}. This is however not surprising, since the VRES matrix
elements are only LO. Including the VRES parts in NLO reduces the
scale dependence, as we have seen above, for bins a and b, i.e., for 
$Q^2\simlt\frac15\mu^2$. We have also looked at scale variations of
cross sections with larger minimum transverse energy. We find the
overall improvement of including a resolved virtual photon component
increasingly smaller, the larger the $E_T$'s are. The upper bound
$Q^2\simlt\frac15\mu^2$ for the improvement of the scale dependence
is also found for these larger $E_T$'s. As an
important result, we found no improvement for $Q^2>10$~GeV$^2$, no
matter how large the $E_T$ was. This is also supported by the
$Q^2$-independent small $K$ factor for $E_T=25$~GeV in
Fig.~\ref{f1}. These findings confirm the expectations from \cite{p2}
that for large enough photon virtualities one should end up with a
fully perturbative prediction irrespective of the probing scale.

The bottom line from these observations is that the resolved virtual
photon approach improves the scale dependence und thus the
perturbative stability of the jet cross sections for $Q^2<10$~GeV$^2$
as long as $Q^2\simlt \frac15 E_T^2$. It is interesting to see that
this result, based on perturbative calculations in NLO QCD and without
making use of the GRS PDF's, agrees with the restrictions implemented
by the GRS group into their old parametrizations of the virtual photon
parton densities \cite{10}. Also for the new PDF's \cite{10b}, for
which no restrictions have been implemented, the authors stress that
the resolved photon approach is only meaningful for $Q^2\ll\mu^2$,
typically $Q^2\simlt \frac{1}{10}\mu^2$. Our results should
have some relevance also for jet production at HERA, since the SR and
DR contributions, which occur in $eP$-scattering, are included in
this analysis and give considerable contributions. We have checked
that the limitations we give above on the $Q^2$-range for the
improvement of the perturbative stability also hold for the SR and DR
contributions alone. However, the perturbative stability in the
$eP$-scattering case should be checked in more detail under HERA
conditions.

A point one has to keep in mind in this discussion is that the
resolved, especially the DR, matrix elements may give important
contributions to jet cross sections,
even though the non-perturbative input from the virtual photon PDF is
small. As we have seen in Fig.~\ref{f2}, the sum of NLO VDIR$_S$ and
NLO VRES gives rather large corrections to the NLO VDIR result, even
for $Q^2$, where the photon PDF is given mainly by the pointlike terms.
These NLO VRES corrections convoluted with the leading logarithmic
contribution to the photon to quark splitting is a leading
logarithmic approximation to the full NNLO result with a 
pure direct virtual photon contribution, which is so far not 
available. The effect of higher order terms in the resolved matrix elements
can be much pronounced in specific phase-space regions, as has been 
shown in \cite{kp}. There, the NLO DR matrix elements are needed to
explain the forward jet production cross section at low $x_{Bj}$ in
$eP$-scattering at HERA in the region $Q^2\in [5,100]$~GeV$^2$.

\section{Jet cross sections at LEP2}

\begin{figure}[hhh]
  \unitlength1mm
  \begin{picture}(122,125)
    \put(-4,-5){\epsfig{file=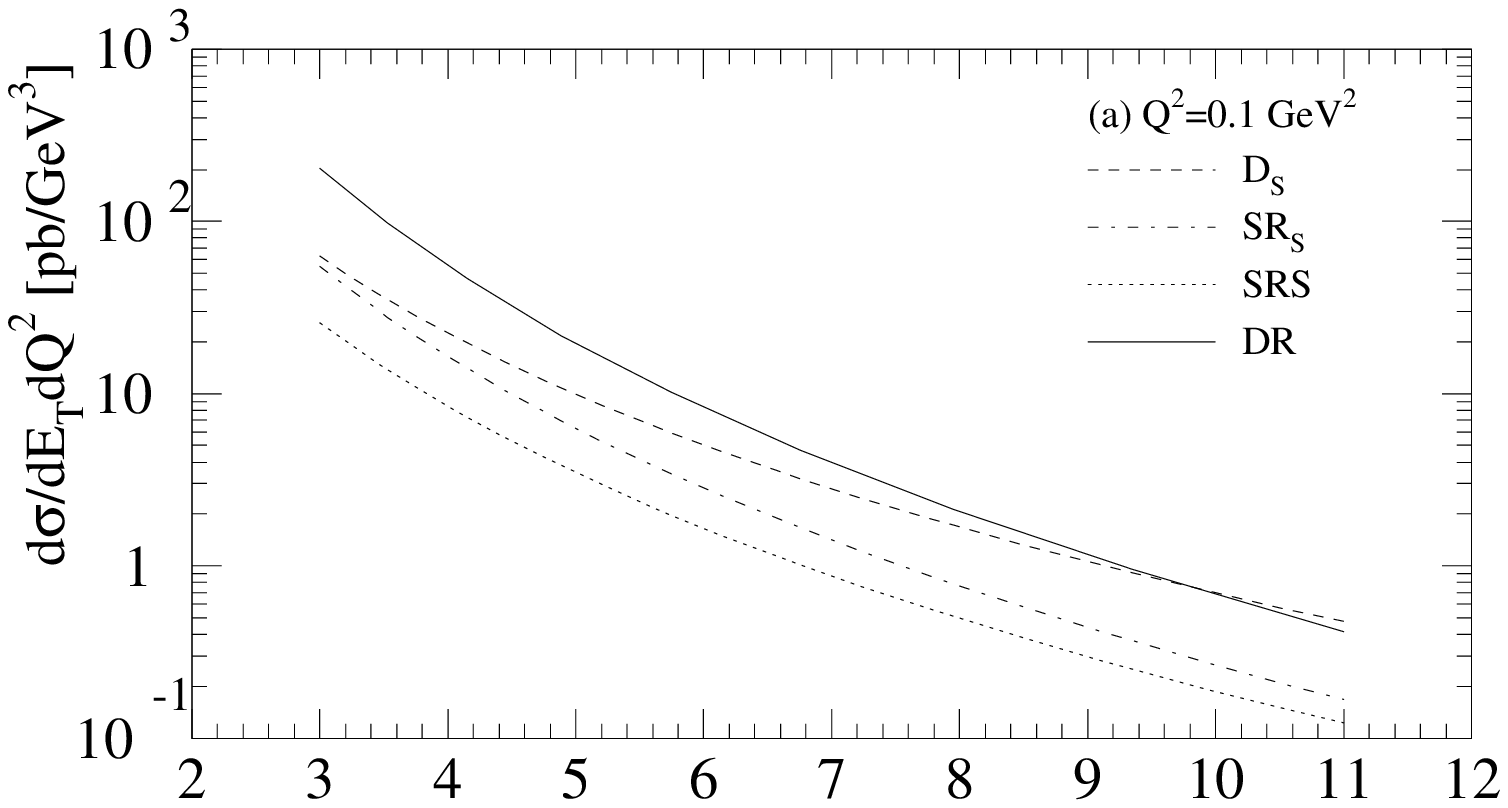,width=9.5cm,height=14cm}}
    \put(78,-5){\epsfig{file=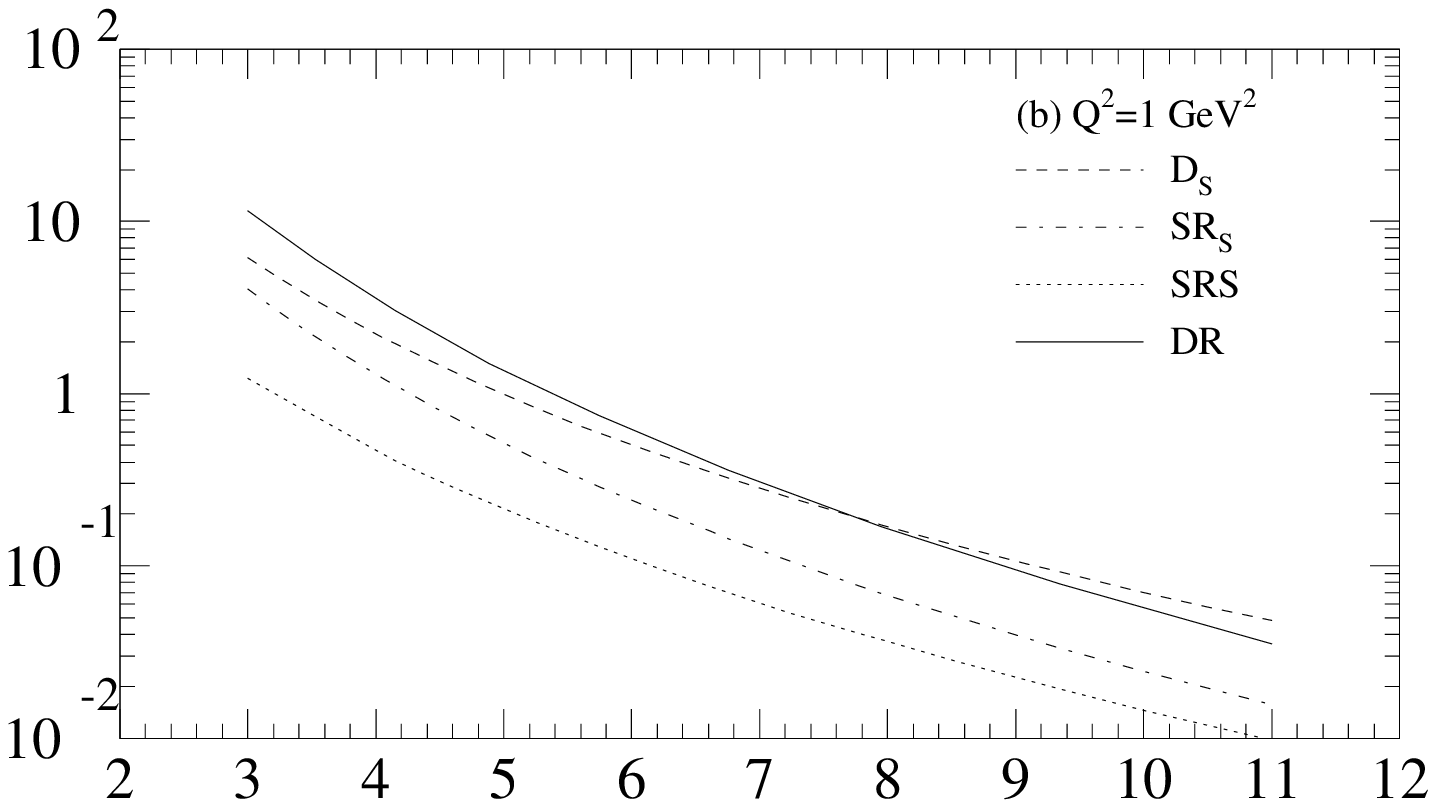,width=9.5cm,height=14cm}}
    \put(-4,-65){\epsfig{file=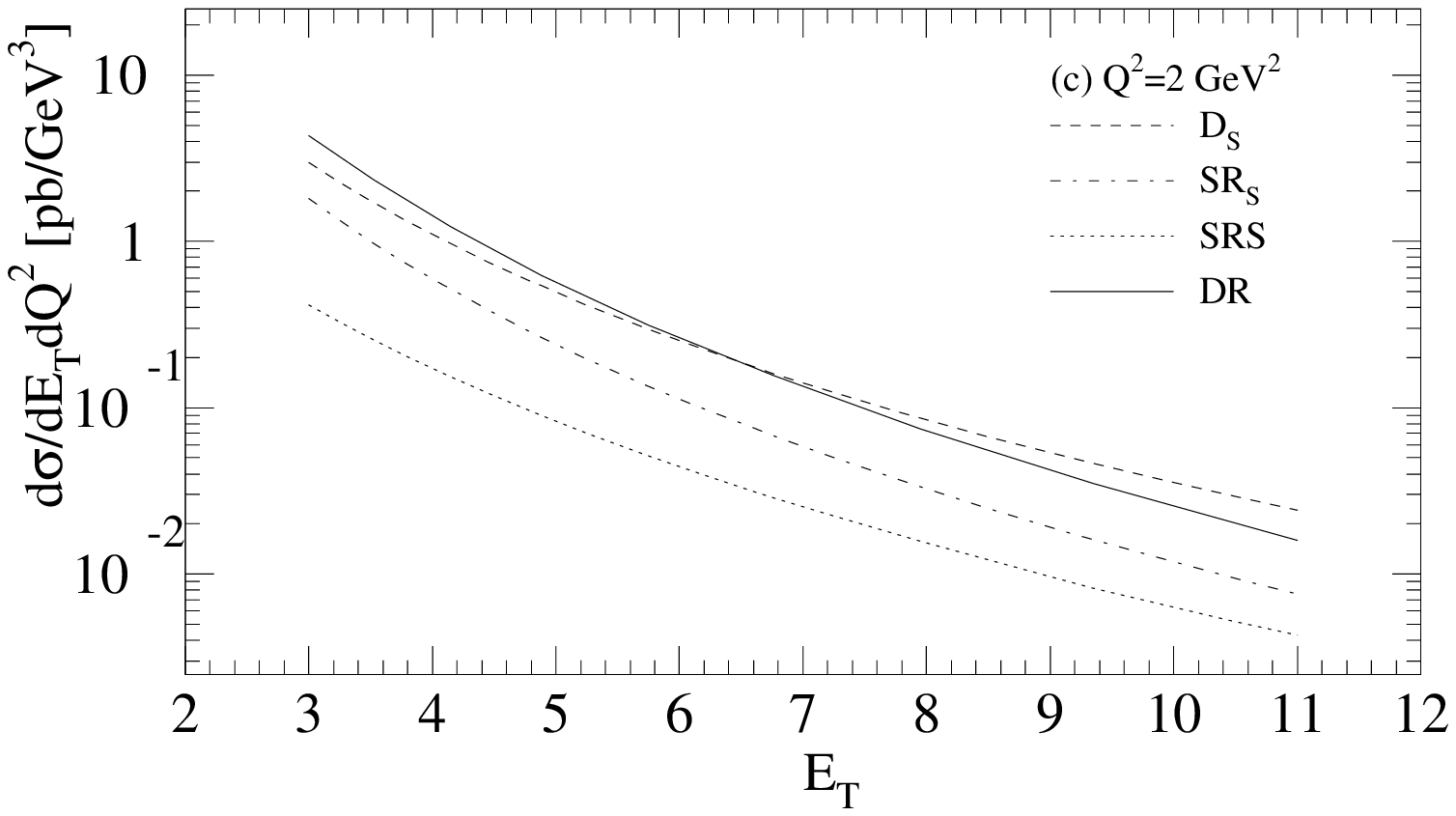,width=9.5cm,height=14cm}}
    \put(78,-65){\epsfig{file=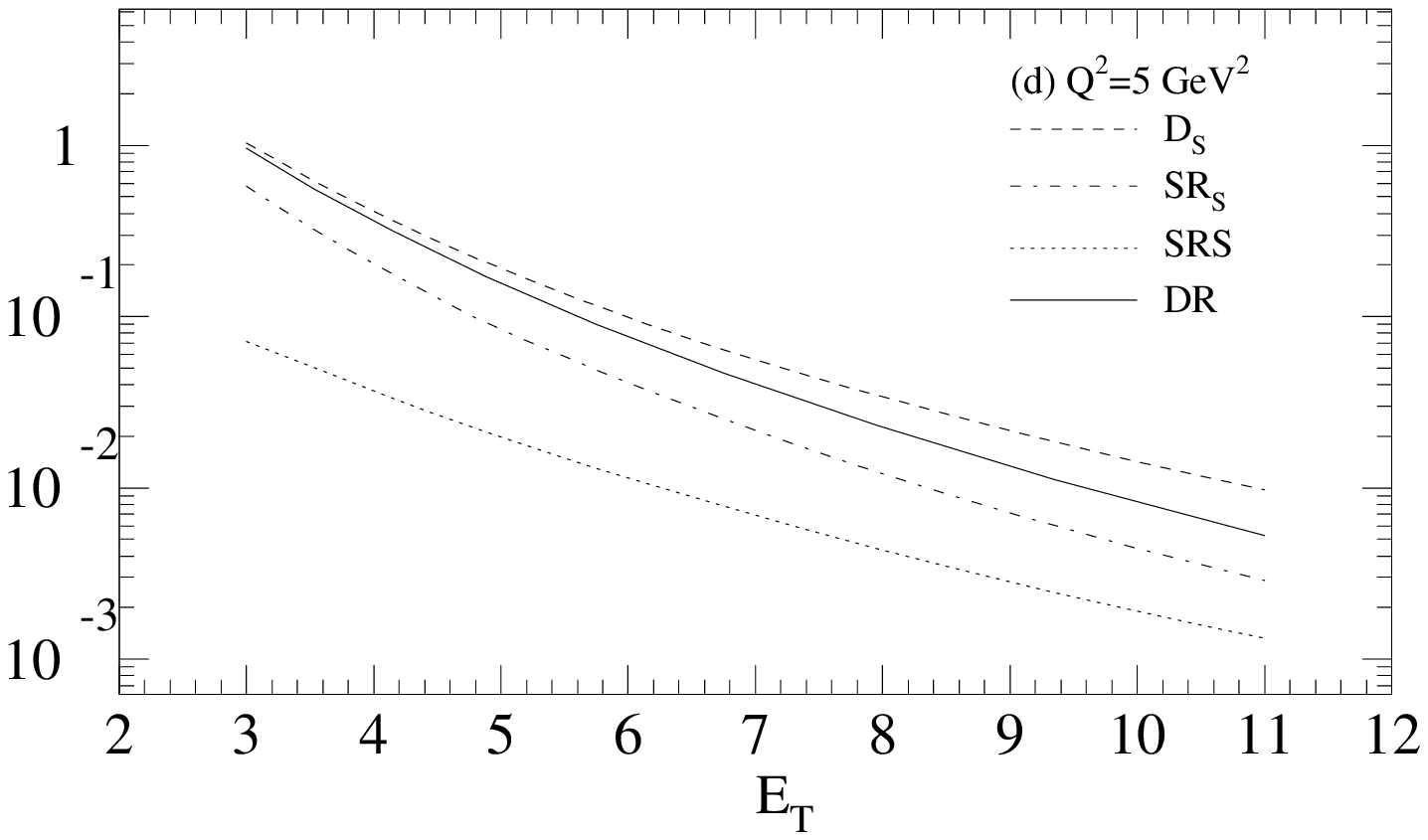,width=9.5cm,height=14cm}}
  \end{picture}
  \caption{\label{f5}\it NLO single jet inclusive cross sections
  integrated over $|\eta|<2$ as functions of $E_T$ for
  different virtualities, employing the SaS PDF's. (a) $Q^2=0.1$~GeV$^2$; 
  (b) $Q^2=1$~GeV$^2$; (c) $Q^2=2$~GeV$^2$;  (d) $Q^2=5$~GeV$^2$.}
\end{figure}

We now turn to predictions of cross sections under conditions to be
met at LEP2. We keep the numerical input as specified at the end of
section 2. We start with absolute predictions for single jet inclusive
cross sections integrated over rapidity $|\eta|<2$
\begin{equation} \label{sjxsec}
  \frac{d\si^{1jet}}{dE_TdQ^2} = \int d\eta
     \frac{d\si^{1jet}}{dE_TdQ^2d\eta} 
\end{equation}
as functions of the transverse energy. In Fig.~\ref{f5}~a--d the
spectra are shown for $Q^2=\frac{1}{10},1,2$ and $5$~GeV$^2$. We have
plotted all four components to see their relative importance. To be
able to include the resolved contributions we have subtracted the
logarithmic terms in the direct components, leading to the D$_S$
(dashed) and SR$_S$ (dash-dotted) curves. One sees the strong fall-off
in $E_T$ and the decrease of the absolute values with increasing
$Q^2$. For $Q^2=0.1$~GeV$^2$ the DR component is rather important for
the whole shown $E_T$ region and especially dominates at
$E_T=3$~GeV. This is in accordance with the photoproduction results
presented in Fig.~\ref{f0}~a. Movin towards larger virtualities, the
influence of the DR cross section becomes smaller, and at
$Q^2=5$~GeV$^2$ the D component is the dominant contribution. Only for
the smallest $E_T$ values is the DR cross section in the same
magnitude as the D one. The SR and SRS contributions are always
small. In addition, for the SRS cross section a further suppression can
be seen for increasing $Q^2$ due to the suppression of the virtual
photon PDF. The virtual resolved components have a stronger fall-off with
$E_T$ than the virtual direct contributions and the relative
importance of the VRES cross sections diminish with increasing $Q^2$.

\begin{figure}[hhh]
  \unitlength1mm
  \begin{picture}(122,125)
    \put(-4,-5){\epsfig{file=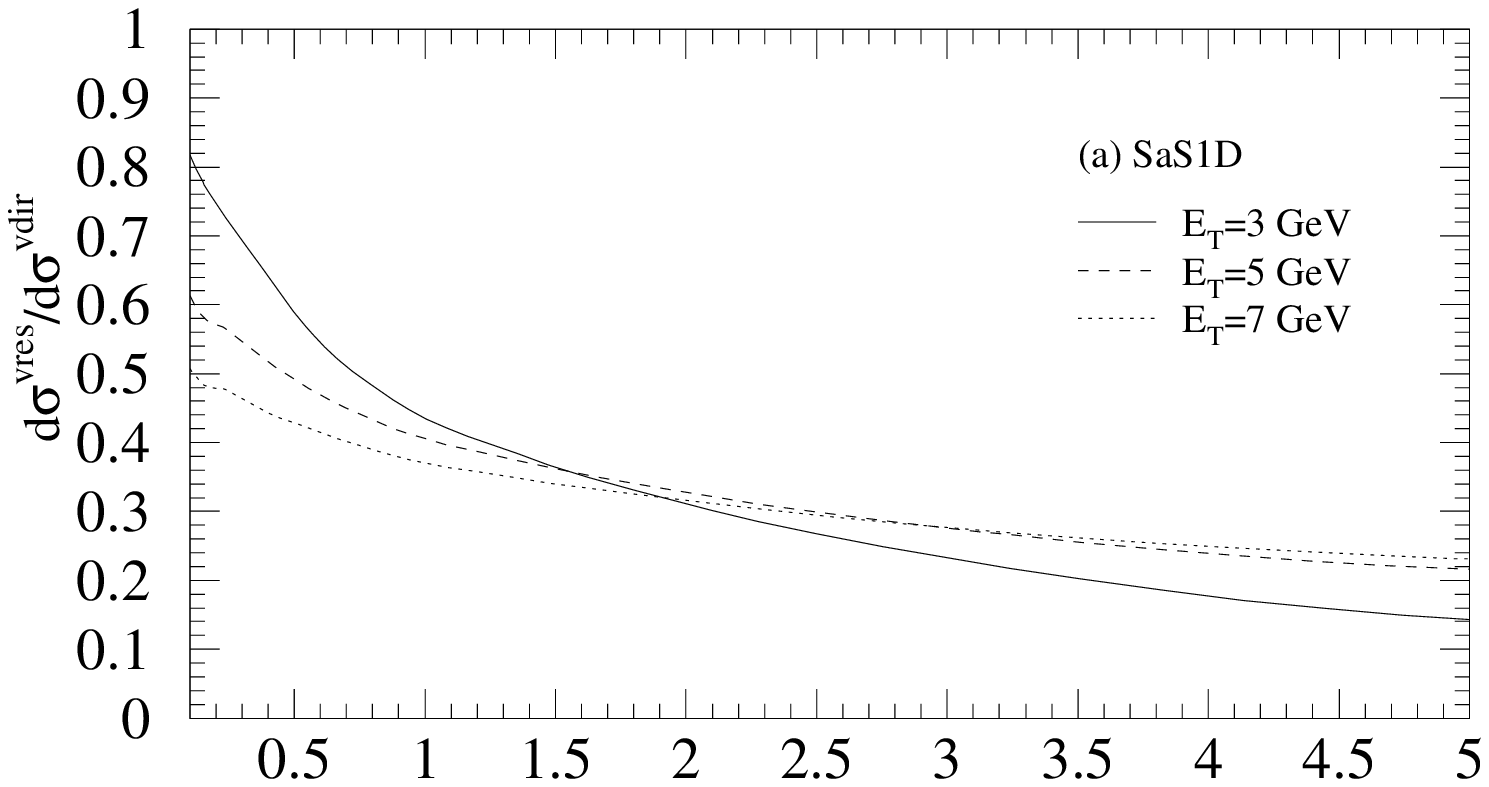,width=9.5cm,height=14cm}}
    \put(78,-5){\epsfig{file=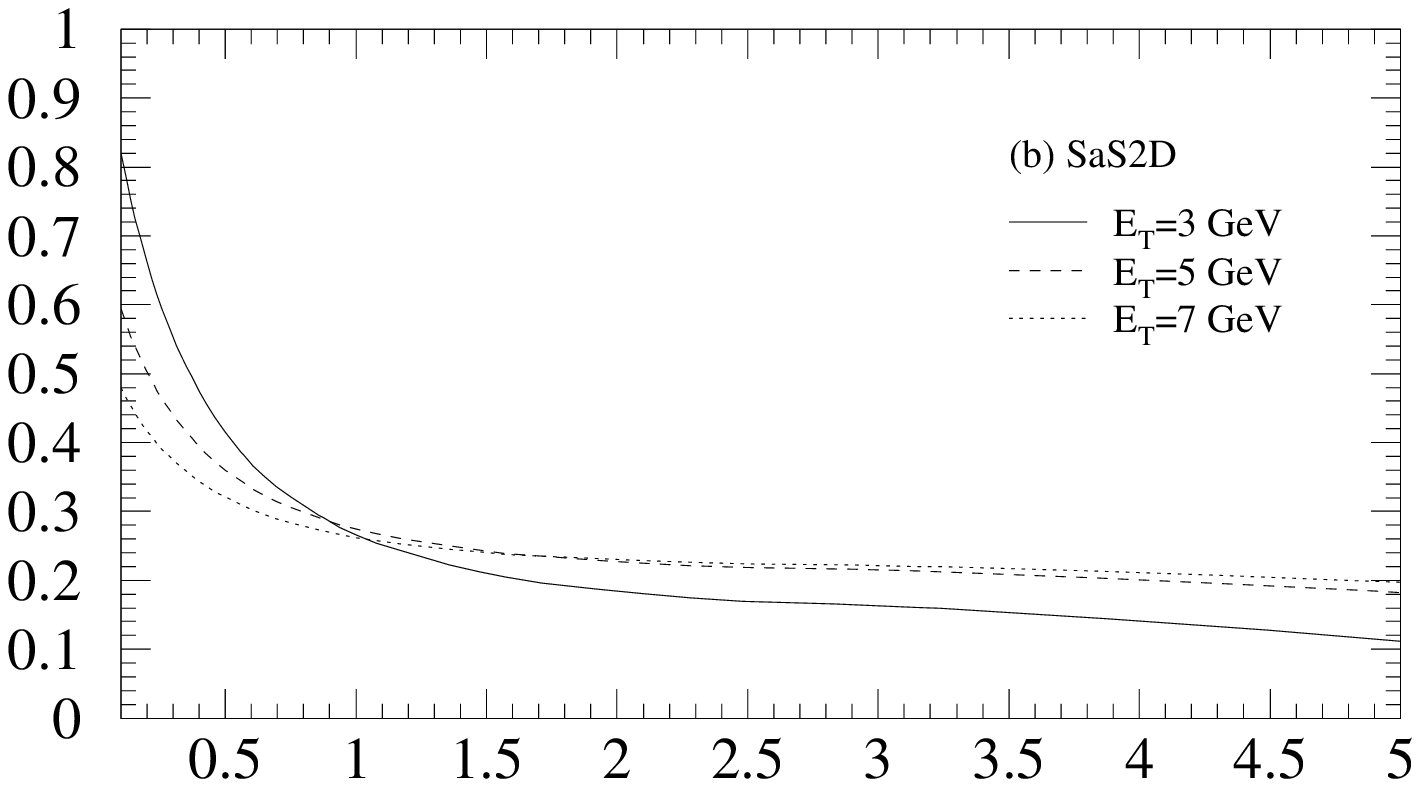,width=9.5cm,height=14cm}}
    \put(-4,-65){\epsfig{file=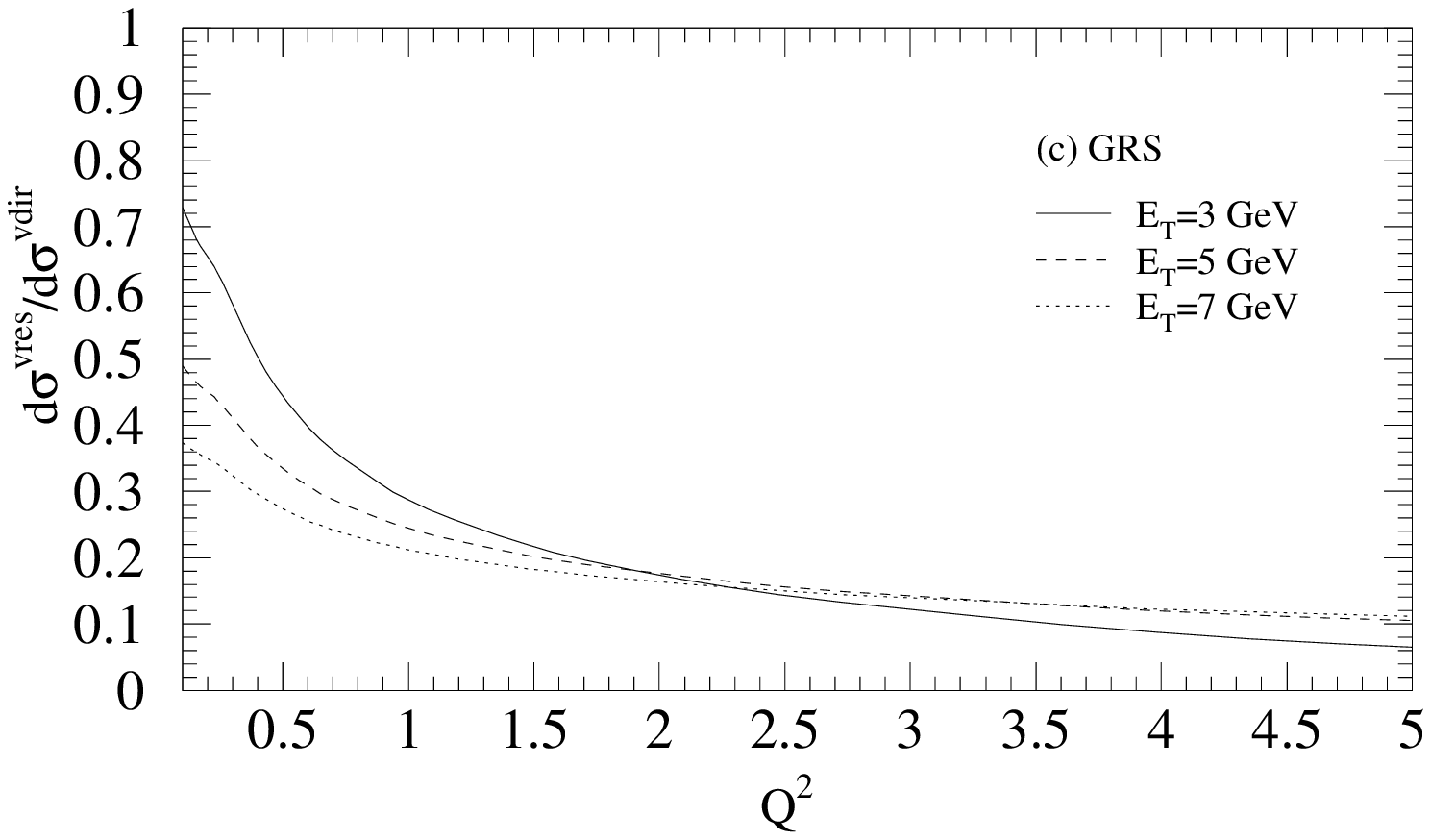,width=9.5cm,height=14cm}}
    \put(78,-65){\epsfig{file=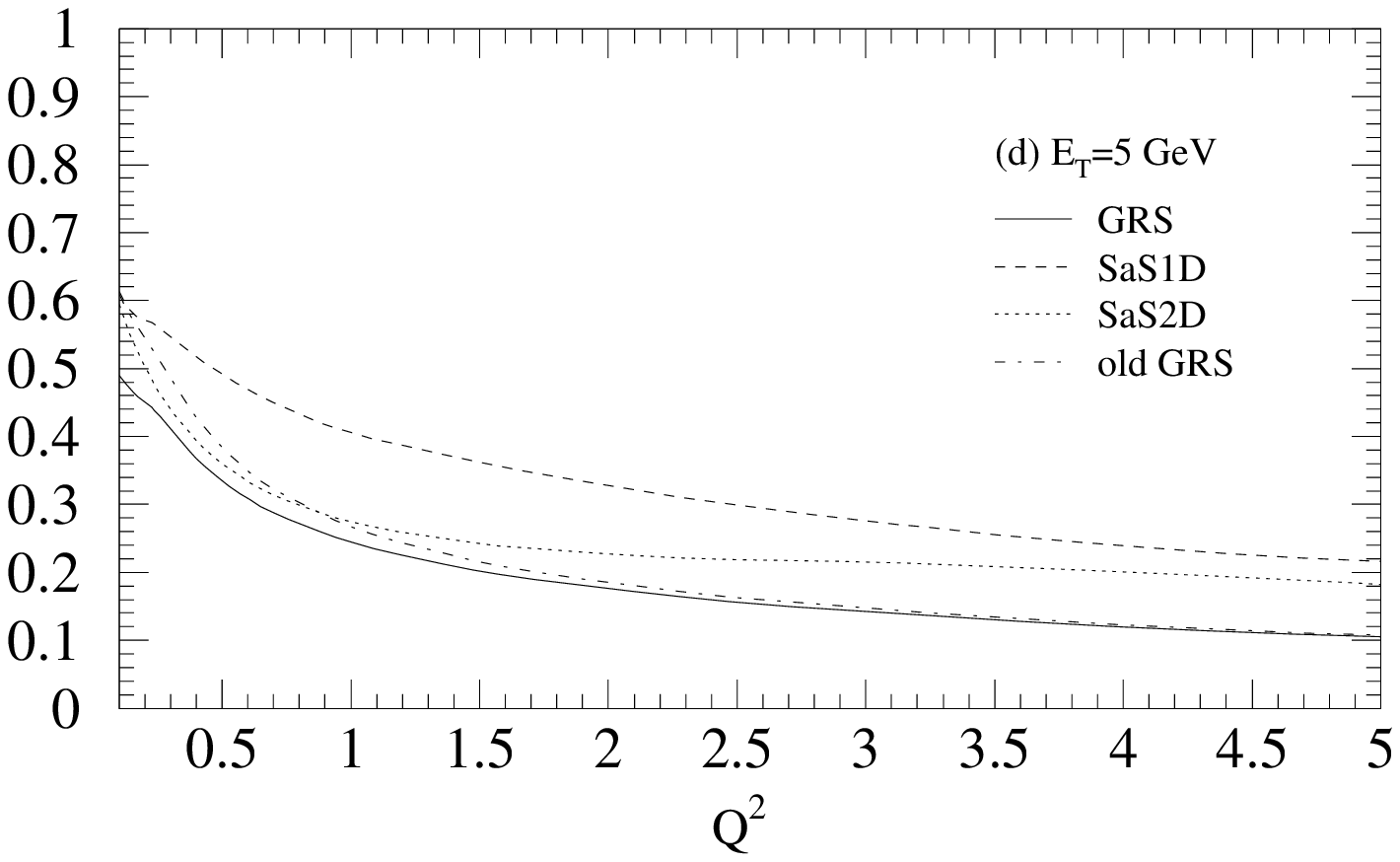,width=9.5cm,height=14cm}}
  \end{picture}
  \caption{\label{f4}\it LO ratio of VRES over VDIR of one-jet inclusive
  cross sections, integrated over $|\eta|<2$ for three different
  scales $E_T=3,5$ and $7$~GeV for different virtual photon PDF's. 
  (a) SaS1D; (b) SaS2D; (c) GRS; (d) comparing the PDF's from a--c at
  $E_T=5$~GeV together with the old GRS PDF.}
\end{figure}

To see in more detail how large the fraction of resolved contributions
in the jet cross sections are, we have calculated the ratio of the
VRES over the VDIR one-jet inclusive cross sections in LO integrated
over $|\eta|<2$ for three different values $E_T=3,5$ and $7$~GeV
for different virtual photon PDF's as functions of $Q^2$ in the region
$Q^2\in [0.1,5]$~GeV$^2$. Note, that in the LO case no subtraction of
the logarithmic terms has to be performed. As a scale we have chosen
$\mu_R=\mu_F=E_T$. The results are shown in Fig.~\ref{f4}~a--d. For all
three PDF's in Fig.~\ref{f4}~a--c one observes the expected fall off
with rising $Q^2$. The fall off is stronger for the smaller scales. For
the larger scales the VRES component diminishes with respect to the
VDIR component, but reaches out farther into the larger $Q^2$
region. For the SaS parametrizations the ratio is around $0.8$ below
$Q^2=0.5$~GeV$^2$ for $E_T=3$~GeV and falls off to $0.5$ for
$E_T=7$~GeV. At $Q^2=5$~GeV the VRES cross section gives about
15--25\% of the VDIR one. These results hold also for the SaS2D case,
shown in Fig.~\ref{f4}~b, although the SaS1D and SaS2D parametrizations
have rather different $Q^2$-behaviour due to the different evolution
starting scales of $Q_0=0.6$~GeV and $Q_0=2$~GeV, respectively. The
SaS2D decreases stronger  and is flatter for larger $Q^2$ than the
SaS1D PDF.  This can also be seen in Fig.~\ref{f4}~d for the scale 
$E_T=5$~GeV, where the two SaS parametrizations can be directly
compared as the dashed and the dotted lines. In the same figure also
the GRS parametrization \cite{10b} (full line) is shown, which
produces a still different fall-off behaviour. For the smaller $Q^2$
the decrease is similar to the SaS2D case but continues stronger for
the larger $Q^2$. This behaviour is seen for the GRS PDF in
Fig.~\ref{f4}~c for all three scales. For comparison we have plotted
in Fig~\ref{f4}~d also the old version of the GRS PDF \cite{10}
(dash-dotted line), which is rather similar to the new
parametrization. It is clear from these curves, especially
Fig.~\ref{f4}~d, that the $Q^2$ dependence is rather different for the
GRS and SaS parametrizations which reflects the ambiguities due to
the limited data on the virtual photon structure function. 

We could repeat the calculation of these ratios for the NLO case, but
the VDIR and VRES components alone depend rather strongly on $\mu_F$. Only
their sum is independent from $\mu_F$. It is preferable to adopt the
strategy used experimentally to distinguish direct and resolved cross
sections for dijet events. The OPAL collaboration have presented
distributions in $x_\gamma^\pm$ for $\gamma\gamma$-scattering
\cite{opal}, where 
\begin{equation} \label{obs}
  x_\gamma^{\pm} = \frac{\sum_{jets}(E\pm p_z)}{\sum_{hadrons} (E\pm
  p_z)} . \
\end{equation}
The sum in the numerator runs over the two largest $E_T$ jets in the
event. The direct dominated cross sections then
corresponds to those contributions, where $x_\gamma^\pm>0.8$ and the
resolved dominated correspond to $x_\gamma^\pm<0.8$. The comparisons
with NLO calculations show good agreement \cite{kkk}. We however do
not show curves like this here since it is not clear whether the
statistics in the $\gamma^*\gamma$-case will be high enough to extract
such distributions. 

\begin{figure}[hhh]
  \unitlength1mm
  \begin{picture}(122,125)
    \put(-4,-5){\epsfig{file=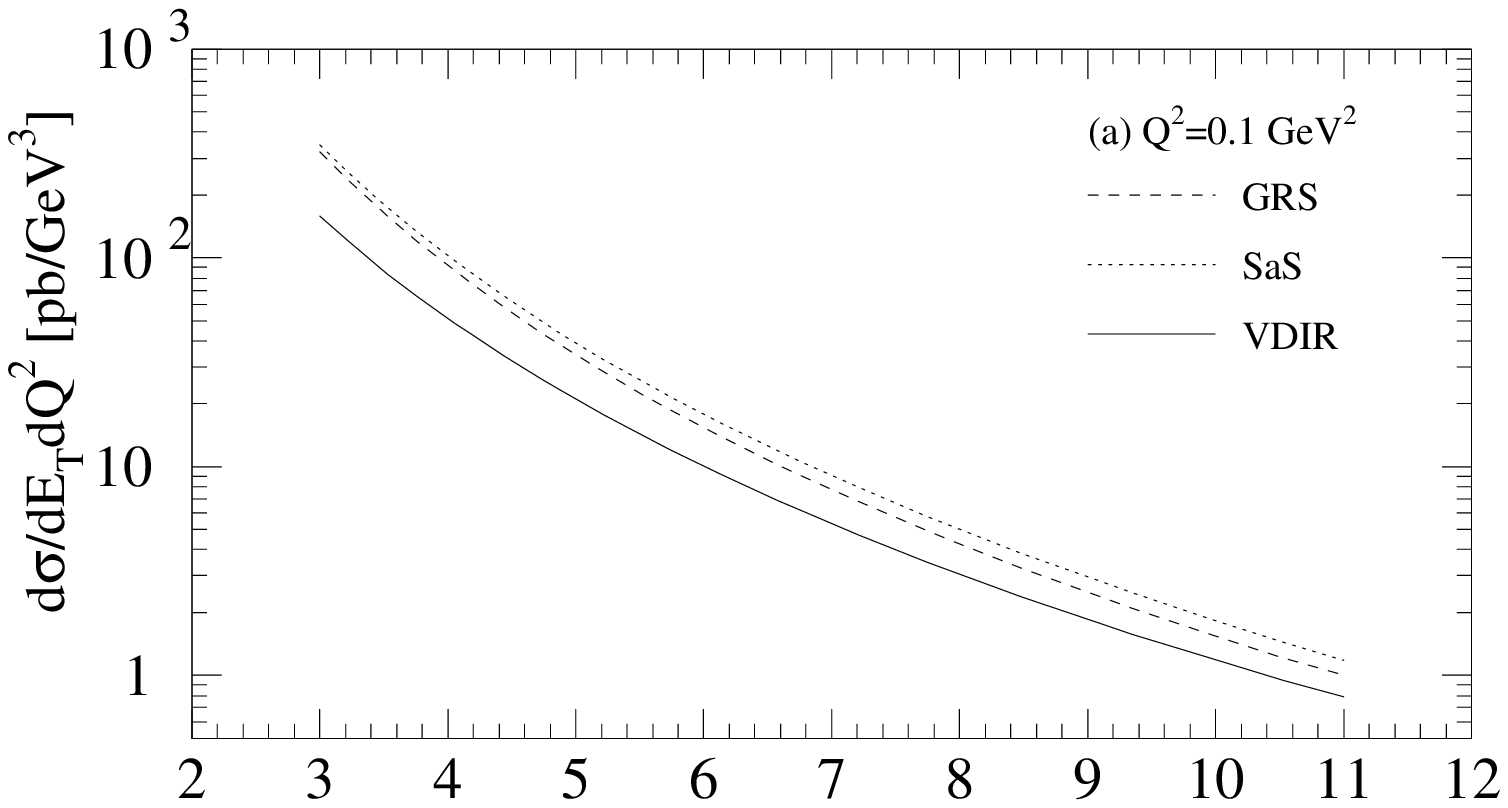,width=9.5cm,height=14cm}}
    \put(78,-5){\epsfig{file=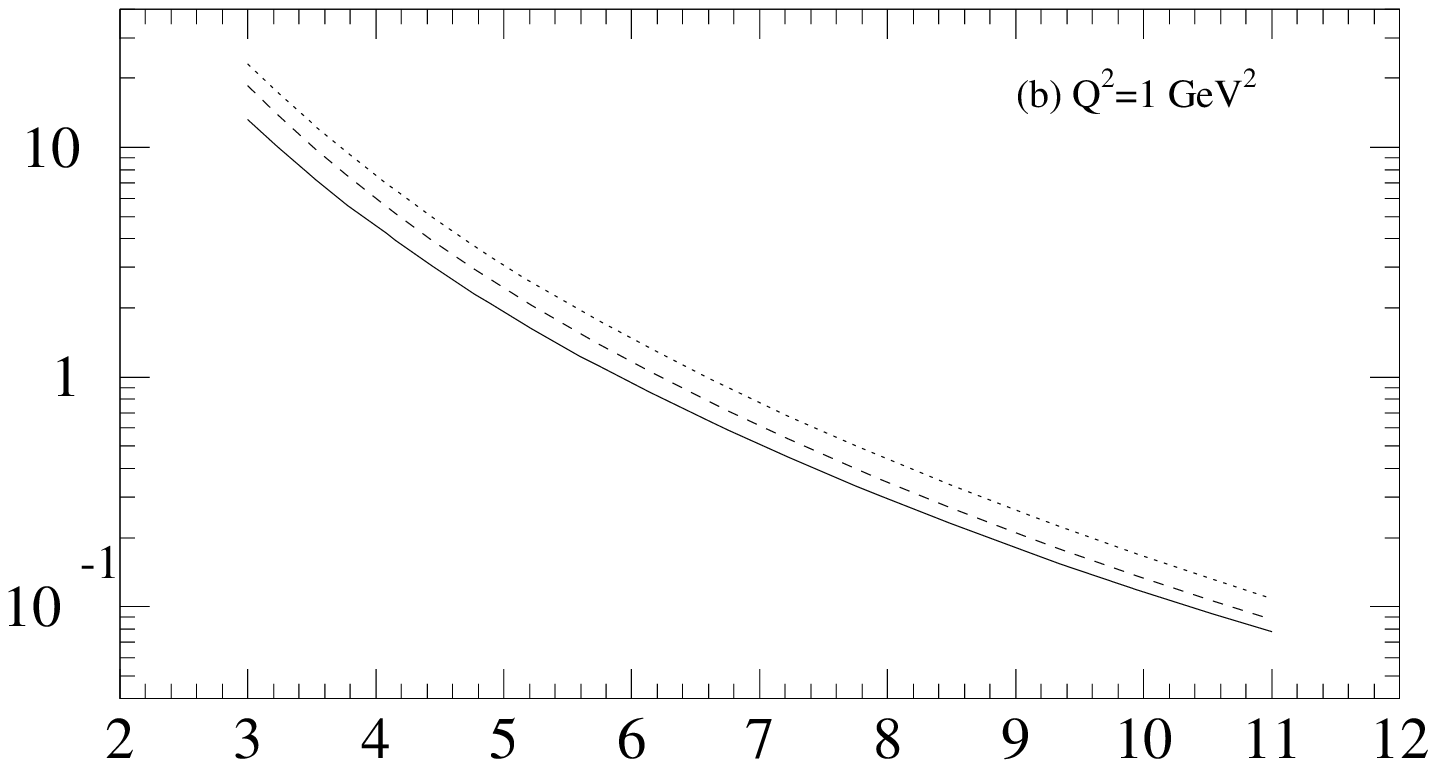,width=9.5cm,height=14cm}}
    \put(-4,-65){\epsfig{file=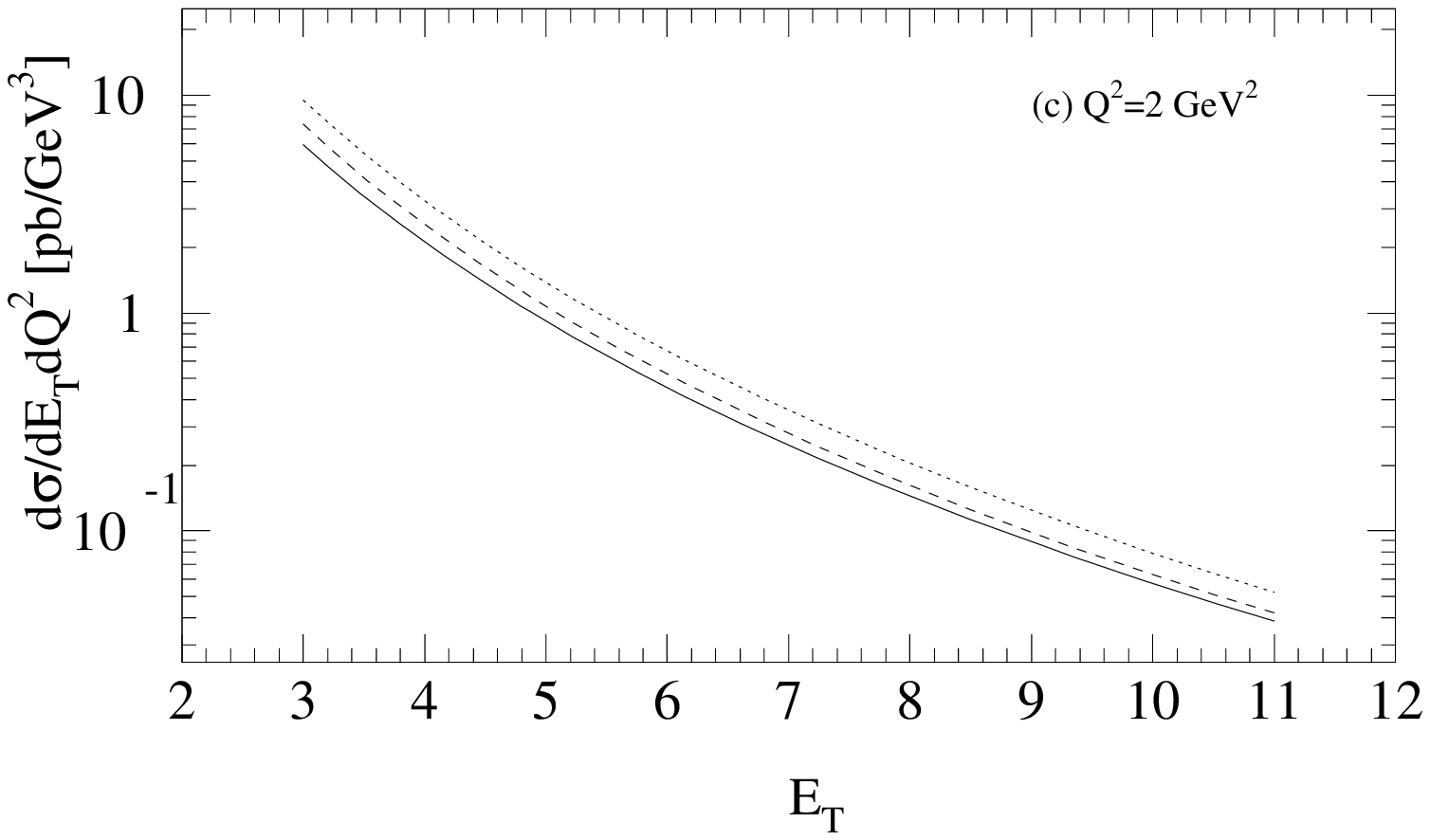,width=9.5cm,height=14cm}}
    \put(78,-65){\epsfig{file=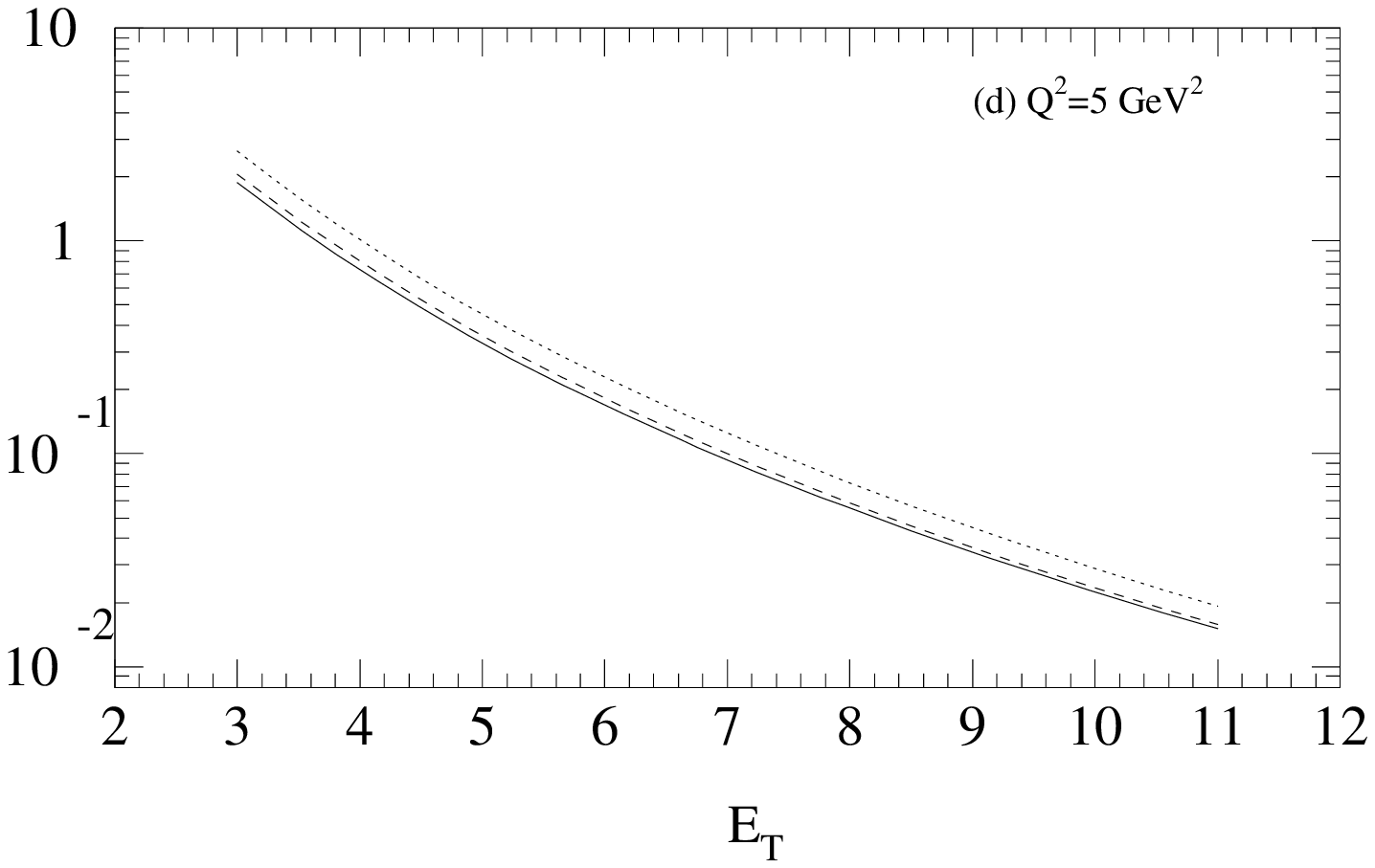,width=9.5cm,height=14cm}}
  \end{picture}
  \caption{\label{f6}\it NLO single jet inclusive cross sections
  integrated over $|\eta|<2$ as functions of $E_T$ for $Q^2$
  values as in Fig.~\ref{f5}~a--d; the full line gives the
  unsubtracted direct (DIS) and has to be compared with the curves
  denoted GRS and SaS. These are the sum of NLO subtracted direct and
  NLO resolved cross sections, employing the GRS (dashed) or the SaS
  (dotted) parton densities.}
\end{figure}

Instead, we compare absolute single-jet cross sections like
(\ref{sjxsec}) for the pure VDIR and the VDIR$_S$+RES approaches with
different virtual photon PDF's GRS and SaS, which are shown in
Fig.~\ref{f6}~a--d. For $Q^2=0.1$~GeV$^2$ the VRES approach yields
about a factor of 2 larger cross sections. This is expected, since
this small virtuality lies in the photoproduction domain. The
predictions from the SaS and the GRS PDF's give similar
results. However, for $Q^2=1$~GeV$^2$ these two parametrizations
already give rather different predictions. The GRS curve is for the
smaller $E_T$'s around 50\% larger than the VDIR, 
whereas the SaS is 70\% larger. Furthermore, the SaS curves do not fall
off that strong with increasing $E_T$. In Fig.~\ref{f6}~c, the
difference between the two VRES predictions is quite pronounced. At
$Q^2=5$~GeV$^2$, the SaS curve gives still a 30--40\% larger cross
section than the VDIR, whereas the GRS curve yields basically the same
result as the VDIR. In general, the difference between the VRES and
the VDIR cross sections in the low $Q^2$ region is large enough that
it should be possible to distinguish between these two approaches
experimentally. Furthermore, the difference between the VRES curves
with the SaS and GRS parametrizations is rather large and it should
therefore also be possible to distinguish between these specific
PDF's and to constrain the $Q^2$-dependence of the virtual photon PDF.

\section{Summary}

We have presented a calculation of jet cross sections in
$\gamma^*\gamma$ scattering from $e^+e^-$-collisions at low $Q^2$ in
NLO QCD, employing the phase-space slicing method to extract
singularities in the real corrections. Logarithmic contributions from
the virtual photon initial state have been subtracted and absorbed
into the resolved virtual photon structure function. Comparison with
existing photoproduction calculations in the limit $Q^2\to 0$ showed
very good agreement. 

We have studied the perturbative stability of two approaches to low
$Q^2$ jet production, namely the pure direct coupling of virtual
photons and, secondly, the inclusion of resolved virtual photons by
looking at $K$ factors and scale dependences. We found the resolved
virtual photon approach to improve the perturbative stability for
$Q^2<10$~GeV$^2$ with $Q^2\simlt \frac15 E_T^2$. However, the NLO
corrections to the resolved matrix elements may be important also for
larger $Q^2$.

We further made predictions for inclusive jet production at LEP2 and
found that it should be suitable to experimentally distinguish between
different parametrizations of the virtual photon structure functions,
since the DR contributions are rather important at low $Q^2$. It
should also be possible to restrict the $Q^2$ dependence of the
parametrizations by comparing the data to the predictions for $E_T$
spectra of the jet cross sections.


\newpage 

\newpage 

\newpage

\newpage 

\newpage 

\newpage 

\begin{thebibliography}{99}

\bibitem{1}
M. Gl\"uck, E. Reya, M. Stratmann, Phys. Rev. D54 (1996) 5515; \\
D. de Florian, C. Garcia Canal, R. Sassot, Z. Phys. C75 (1997) 265; \\
J. Chyla, J. Cvach, Proceedings of the Workshop 1995/96 on "Future
Physics at HERA", eds. G. Ingelman, A. de Roeck and R. Klanner,
DESY 1996, Vol. 1, p. 545

\bibitem{2}
M. Klasen, G. Kramer, B. P\"otter, Eur. Phys. J. C1 (1998) 261; \\
B. P\"otter, DESY-97-138, hep-ph/9707319

\bibitem{3}
G.~Kramer, B.~P\"otter Eur. Phys. J. C5 (1998); see also
\cite{lund98}, p.~29, hep-ph/9810450

\bibitem{lund98}
Proceedings of the Workshop on "Photon interactions and the photon
structure", Lund, Sweden, 10-12 September 1998, eds.\ G.~Jarlskog and
T.~Sj\"ostrand

\bibitem{4}
 H1 Collaboration (C.~Adloff et al.), Phys. Lett. B415 (1997) 418;
 DESY-98-076, hep-ex/9806029;  DESY-98-205, hep-ex/9812024; \\
 ZEUS Collaboration (J.~Breitweg et al.), 29th International
 Conference on High Energy Physics (ICHEP98), Vancouver, Canada, 23-29
 July 1998

\bibitem{9} PLUTO Collaboration (Ch.~Berger et al.), Phys. Lett.
    B142 (1984) 119

\bibitem{10} M.~Gl\"uck, E.~Reya, M.~Stratmann, Phys. Rev. D51
  (1995) 3220

\bibitem{10b} M.~Gl\"uck, E.~Reya, I.~Schienbein, DO-TH 99/03,
  February 1999, hep-ph/9903337

\bibitem{11} G.A.~Schuler, T.~Sj\"ostrand, Z. Phys. C68 (1995) 607;
  Phys. Lett. B376 (1996) 193

\bibitem{opal2} OPAL Collaboration, OPAL physics note PN293, May
1997. Contribution to XVIII Intern. Symposium on Lepton-Photon
Interactions, Hamburg 1997 and to the Intern.\ Europhysics Conference
om HEP, Jerusalem 1997

\bibitem{12} B.~P\"otter, Nucl.~Phys.~B540 (1999) 382

\bibitem{kk} M.~Klasen, G.~Kramer, Z.~Phys.~C72 (1996) 107

\bibitem{kkk} M.~Klasen, T.~Kleinwort, G.~Kramer, EPJ direct 1 (1998) 1

\bibitem{15} T. Kleinwort, G. Kramer, Nucl. Phys. B477 (1996) 3; 
 Phys. Lett. B370 (1996) 141; Z. Phys. C75 (1997) 489;
 T. Kleinwort, DESY-96-165

\bibitem{14} P. Aurenche, J.-Ph. Guillet, M. Fontannaz, Y. Shimizu,
  J. Fujimoto, K. Kato, Progr. Theor. Phys. 92 (1994) 175; \\
  L.E. Gordon, Nucl. Phys. B419 (1994) 25

\bibitem{p2} T.~Uematsu, T.F.~Walsh, Phys.~Lett.\ B101 (1981) 263,
Nucl. Phys. B199 (1982) 93; \\
G.~Rossi, Phys.~Rev. D29 (1984) 852

\bibitem{13} M.~Stratmann, contribution to \cite{lund98}, p.~183,
hep-ph/9811260

\bibitem{12a} B.P\"otter, contribution to \cite{lund98}, p.~175,
hep-ph/9810466; Proceedings of International Euroconference on
Quantum Chromodynamics (QCD 98), Montpellier, 
France, 2-8 July 1998, hep-ph/9807538

\bibitem{wwill} C.F. v. Weizs\"acker, Z. Phys. 88 (1934) 612; 
  E.J. Williams, Kgl. Danske Vidensk. Selskab. Mat-Fiz. Medd. 13
  (1935) N4

\bibitem{schuler} G.A.~Schuler, CERN-TH/96-297, October 1996,
  Phys. Lett. B (in press)

\bibitem{graud} D. Graudenz, Phys. Rev. D49 (1994) 3291

\bibitem{jv} B.~P\"otter, DESY-98-071, hep-ph/9806437,
 Comp.~Phys.~Comm. (in press) 

\bibitem{grv} M. Gl\"uck, E. Reya, A. Vogt, Phys. Rev. D45 (1992) 
  3986; Phys. Rev. D46 (1992) 1973

\bibitem{snow} J.E. Huth et al., Proc. of the 1990 DPF Summer Study
 on High Energy Physics, Snowmass, Colorado, edited by E.L. Berger, World
 Scientific, Singapore, 1992, p. 134

\bibitem{kp} G.~Kramer, B.~P\"otter, DESY-99-004, MPI--PhT 99-01,
hep-ph/9901314

\bibitem{opal} OPAL Collaboration, OPAL physics note 291, May
1997. Contribution to XVIII Intern. Symposium on Lepton-Photon
Interactions, Hamburg 1997; Proceedings of the Ringberg Workshop on
"New Trends in HERA Physics", Ringberg 1997.

\end{thebibliography}
\end{document}